\documentclass[11pt,a4paper]{article}

\usepackage[a4paper,text={150mm,240mm},centering,headsep=10mm,footskip=15mm]{geometry}
\usepackage[T1]{fontenc}
\usepackage{epsfig, float,color}
\usepackage{amsmath,amssymb}
\usepackage{amsfonts}
\usepackage{url}
\usepackage{bbm}
\usepackage{mathrsfs} 
\usepackage{cite}

\newcommand{\D}{\dagger}

\newcommand{\R}{\mathbb{R}}
\newcommand{\C}{\mathbb{C}}

\newcommand{\id}{\mathbbm{1}}

\newcommand{\diag}{{\rm diag} }

\newenvironment{proof}[1][Proof:]{\begin{trivlist}
\item[\hskip \labelsep {\bfseries #1}]}{\end{trivlist}}

\newenvironment{remark}[1][Remark:]{\begin{trivlist}
\item[\hskip \labelsep {\bfseries #1}]}{\end{trivlist}}

\newcommand{\qed}{\nobreak \ifvmode \relax \else
      \ifdim\lastskip<1.5em \hskip-\lastskip
      \hskip1.5em plus0em minus0.5em \fi \nobreak
      \vrule height0.75em width0.5em depth0.25em\fi}

\title{On the question of current conservation for the Two-Body Dirac equations of constraint theory}

\author{
Matthias Lienert\thanks{lienert@math.lmu.de, Mathematisches Institut, Ludwig-Maximilians-Universit\"at, 
	Theresienstr.\@ 39, 80333 M\"unchen, Germany}
}

\date{March 9, 2015}

\begin{document}

\maketitle

\begin{abstract}
	\noindent The Two-Body Dirac equations of constraint theory are of special interest not only in view of applications for phenomenological calculations of mesonic spectra but also because they avoid no-go theorems about relativistic interactions. Furthermore, they provide a quantum mechanical description in a manifestly Lorentz invariant way using the concept of a multi-time wave function. In this paper, we place them into the context of the multi-time formalism of Dirac, Tomonaga and Schwinger for the first time. A general physical and mathematical framework is outlined and the mechanism which permits relativistic interaction is identified. The main requirement derived from the general framework is the existence of conserved tensor currents with a positive component which can play the role of a probability density. We analyze this question for a general class of Two-Body Dirac equations thoroughly and comprehensively. While the free Dirac current is not conserved, it is possible to 
 find replacements. 
Improving on 
        previous research, we achieve definite conclusions whether restrictions of the function space or of the interaction terms can guarantee the positive definiteness of the currents -- and whether such restrictions are physically adequate. The consequences of the results are drawn, with respect to both applied and foundational perspectives.\\
    
\end{abstract}

\section{Introduction}

In view of the many open questions in quantum field theory (QFT) of both conceptional and mathematical nature, most prominently the measurement problem and the ultraviolet (UV) divergencies \cite{bell_trieste}, it seems a worthwhile question to ask whether there exists an intermediary level between non-relativistic quantum mechanics (QM) and QFT: a relativistic $N$-particle theory.\\
In this regard, the development in constraint theory \cite{dirac_lectures} has led to the formulation of the \textit{Two-Body Dirac} (2BD) \textit{equations} \cite{2bd,sazdjian_2bd,2bd_scalar_vector,2bdem,2bd_hyperbolic,mourad_sazdjian,crater_applications}. These are simultaneous relativistic equations for a 16-component wave function $\psi(x_1,x_2)$ which relates the space-time coordinates $x_i$ of two particles $i = 1,2$. They take the following form:
\begin{align}
 D_1 \psi(x_1,x_2)~&=~0, \nonumber\\
 D_2 \psi(x_1,x_2)~&=~0,
 \label{eq:2bdgeneralform}
\end{align}
where $D_i ~=~ D_{i,0} + \tilde{V}_i(x,\hat{P})$. Here, $D_{i,0}$ is the operator in the free Dirac equation $D_{i,0} \psi = 0$ for the $i$-th particle in manifestly covariant form and $\tilde{V}_i(x,\hat{P})$ are functions of $x = x_1-x_2$ and the total momentum operator $\hat{P}$.\\
The 2BD equations are interesting with respect to several aspects:
\begin{enumerate}
 \item They are closed equations for bound state problems.
 \item They provide evolution equations for a multi-time wave function.
 \item They include a new mechanism of relativistic interaction.
\end{enumerate}
Concerning 1., equations of this kind are useful for applications in nuclear and particle physics, such as calculations of meson spectra (see e.g.\@ \cite{crater_applications,crater_quark_model,jallouli_sazdjian_98,crater_meson_spectroscopy}).\\
With regard to point 2, it seems clear that a manifestly covariant description of quantum phenomena in the Schrödinger picture requires the wave function to depend on all of the space-time coordinates $x_i = (t_i,\mathbf{x}_i)$ of each particle $i$. Because of the many time coordinates, these relativistic wave functions are called \textit{multi-time wave functions}. The idea of a multi-time wave function was proposed by Dirac at the very beginning of relativistic quantum theory as one of its key elements \cite{dirac_32,dfp} and it was subsequently developed further by Bloch \cite{bloch}, Tomonaga \cite{tomonaga} and Schwinger \cite{schwinger}. More recently, it has become clear that multi-time wave functions are also of particular interest for the foundations of QM as they allow for an extension of realistic quantum theories to the relativistic domain. Examples include relativistic GRW theories \cite{grwf,rel_grw} and relativistic Bohmian mechanics \cite{hbd,rel_bm}, showing possible
  ways how the measurement 
problem 
can be avoided in a relativistic context.\\
Concerning 3., one may wonder how the 2BD equations escape no-go theorems about direct relativistic interactions such as the ``no interaction theorem'' \cite{nointeraction,nointeraction2} or the recent result \cite{nogo_potentials}. The key feature is the inclusion of ``potential'' terms depending on arbitrary powers of the total momentum operator. Possibilities of this novel -- and perhaps radical kind are left open by the above-mentioned no-go theorems which concern Hamiltonian evolution equations.\\
The inclusion of potential terms of this kind raises several important conceptual questions which have not been clarified fully in the literature on the 2BD equations:
\begin{enumerate}
 \item How should one understand the 2BD equations mathematically if one includes many times and in addition leaves the quantum mechanical Hamiltonian framework?
 \item Does the general framework for a quantum mechanical theory have to be changed because of the fact that the potential terms contain differential operators up to infinite order?
\end{enumerate}
The first question poses itself because of (a) the inclusion of many time variables in a multi-time wave function and (b) the fact that the potential terms in the 2BD equations contain time derivatives of any order. However, not even $-i\tfrac{\partial}{\partial t}$ is an operator on the commonly used Hilbert space $L^2(\mathbb{R}^3)$ as the elements of the latter are functions $f(\mathbf{x})$ of space variables only.\\
The second point concerns the question if the way how a quantum mechanical theory makes predictions is damaged in any way by the unconventional structure of the 2BD equations. In essence, one requires that the wave equations imply the existence of a current and a corresponding density which is positive definite so that it can play the role of a probability density. In non-relativistic QM, the famous Born rule states that $|\psi|^2$ is the correct probability density for the statistical distribution of particles in experiments. In the multi-time picture one seeks a conserved tensor current $j^{\mu \nu}(x_1,x_2)$ (here for two particles), i.e.\@ $\frac{\partial}{\partial x_1^\mu} j^{\mu \nu} = 0 = \frac{\partial}{\partial x_2^\nu} j^{\mu \nu}$. The question then is whether $j^{\mu \nu}$ has a positive component, such as $j^{00} = \psi^\D \psi$ in free two-time Dirac theory. This is not evident anymore for the 2BD equations and was e.g.\@ the subject of \cite{sazdjian_scalarprod}.\\

\noindent In this paper, we approach the above-mentioned questions as follows: We first introduce the perhaps unfamiliar idea of a multi-time wave function and show that it does in fact arise naturally from Lorentz transformations of $N$-particle wave functions from non-relativistic QM (sec.\@ \ref{sec:multitimederiv}). This includes a brief survey of how one can understand time evolution in this framework. We also outline the connection to a Hilbert space setting (sec \ref{sec:multitimecurrents}). Furthermore, we comment on how the 2BD equations evade the no-go theorems for Hamiltonian theories (sec.\@ \ref{sec:nogotheorempotentials}). Then we introduce the mathematical idea underlying the 2BD equations which allows them to satisfy a certain necessary compatibility condition (sec.\@ \ref{sec:implicationskb}). We also discuss its physical consequences, in particular the necessity of the appearance of a certain covariantization of the spatial distance in the center of momentum frame a
 nd the compatibility of 
this fact 
with the theory of relativity. Subsequently, we introduce and focus on a concrete and important class of the 2BD equations for particle-antiparticle pairs which was first suggested by Sazdjian \cite{sazdjian_2bd} (sec.\@ \ref{sec:2bdfermionantifermion}). This class is related to other forms of the 2BD equations like the one used by Crater and Van Alstine (see \cite{2bd_hyperbolic,mourad_sazdjian}) such that our analysis is sufficiently general. We point out open mathematical issues (sec.\@ \ref{sec:basicquestions}) and propose a preliminary understanding (sec.\@ \ref{sec:preliminaryunderstanding}).\\
After this, we come to the main part of this work: the question whether an appropriate tensor current exists. To approach this question, we show that the free Dirac current is not conserved (sec.\@ \ref{sec:previousresults}). Possible replacements can, however, be found. Nevertheless, there does not seem to exist a general argument that any of them is positive definite. We therefore continue the analysis by posing the question whether further conditions can render the currents or equivalently, the associated scalar product, positive definite. A related analysis of Sazdjian \cite{sazdjian_scalarprod} is discussed and found incomplete. The main open points are the questions of 1. how one should regard the further assumptions, (a) as restrictions on the admitted space of functions (sec.\@ \ref{sec:restrictedfunctions}) or (b) as restrictions on the admissible potentials \ref{sec:restrictedpotentials}, as well as 2. whether the two options (a), (b) or a combination thereof is physically 
 reasonable. These 
questions are answered and the consequences are drawn. Furthermore, we briefly point out that the class of gauge transformations for the 2BD equations is restricted as compared with the free case (sec.\@ \ref{sec:gaugeinvariance}). The paper ends with a discussion of the results and their implications.

\section{Multi-time wave functions} \label{sec:multitime}
In this section, we provide some general background for multi-time wave functions and outline their connection with physics. This is necessary for a proper understanding of the 2BD equations as evolution equations for a wave function $\psi(x_1,x_2)$ with $x_i = (t_i,\mathbf{x}_i)$.

\subsection{Derivation, domain and evolution equations} \label{sec:multitimederiv}
First, we show how the idea of a multi-time wave function arises naturally from general considerations. Consider an $N$-particle system with single-time wave function $\varphi(\mathbf{x}_1,...,\mathbf{x}_N,t)$ as a solution of a single-time wave equation such as Schrödinger's  or Breit's equation \cite{breit_eq}. The argument $(\mathbf{x}_1,...,\mathbf{x}_N,t)$ of the wave function describes a configuration of $N$ space points at a certain time $t$. If we want to Lorentz-transform the wave function, we expect from basic representation theory that the argument of the wave function should transform according to the inverse of the Lorenz transformation $\Lambda$. However, in order to have a Lorentz transformation act on the configuration, we need to regard it as physically synonymous to the following collection of four-vectors at equal time: $((t,\mathbf{x}_1),...,(t,\mathbf{x}_N))$.
Now we can act with $\Lambda^{-1}$ as follows: $(\Lambda^{-1}(t,\mathbf{x}_1),...,\Lambda^{-1}(t,\mathbf{x}_N)) \equiv ((t_1',\mathbf{x}_1'),...,(t_N',\mathbf{x}_N'))$ where in general all $t_k'$ are different. In this way, we see that the Lorentz transformation of a single time wave function leads to the necessity to consider \textit{multi-time wave functions}
\begin{equation}
 \psi : \mathscr{S} \subset \underbrace{\R^4 \times \cdots \times \R^4}_{N~{\rm times}} ~\longrightarrow~\mathcal{S},~~~(x_1,..., x_N) ~\longmapsto~\psi (x_1,..., x_N).
 \label{eq:multitimewavefn}
\end{equation}
The connection to single-time wave functions $\varphi$ is given by
\begin{equation}
 \varphi(\mathbf{x}_1,...,\mathbf{x}_N,t) ~=~ \psi(t,\mathbf{x}_1,...,t,\mathbf{x}_N).
 \label{eq:connection}
\end{equation}
Because the Lorentz transformation of a simultaneous configuration yields a space-like configuration, the natural domain $\mathscr{S}$ of $\psi$ is the set of \textit{space-like configurations\footnote{Note that for $N > 2$, this is a slight (and relativistically natural) generalization of what the Lorentz transformations of simultaneous configurations yield.},} i.e.\@
\begin{equation}
 \mathscr{S} ~=~ \{ (x_1,...,x_N) \in \R^{4N} : (x_i-x_j)_\mu(x_i-x_j)^\mu < 0~\forall i \neq j \}.
 \label{eq:omega}
\end{equation}
Our sign convention for the flat-space metric is $g = \diag(1,-1,-1,-1)$.\\
Dirac's original suggestion for multi-time evolution equations was to prescribe $N$ simultaneous systems of first order partial differential equations \cite{dirac_32}:
\begin{align}
 i \frac{\partial}{\partial {t_1}} \psi ~&=~ H_1 \psi\nonumber\\
 &~ \,\vdots\nonumber\\
  i \frac{\partial}{\partial {t_N}} \psi ~&=~ H_N \psi,
  \label{eq:multitime}
\end{align}
where the $H_k,~k=1,...,N$ are differential operators on an appropriate function space and do not contain time derivatives. Dirac's idea was to prescribe eq.\@\@ \eqref{eq:multitime} on $\R^{4N}$ but it became clear shortly after that this is problematic for two reasons \cite{bloch,tomonaga}: Firstly, the dynamics may exist on $\mathscr{S}$ but not on $\R^{4N}$ and secondly, the statistical meaning of the wave function can in general only be valid on space-like hypersurfaces.\\
For a single-time wave function, one usually prescribes initial data $\varphi(\mathbf{x}_1,...,\mathbf{x}_N,0) \stackrel{!}{=} \varphi_0(\mathbf{x}_1,...,\mathbf{x}_N)$. Using the connection between single- and multi-time wave functions \eqref{eq:connection} we translate this into:
\begin{equation}
 \psi(0,\mathbf{x}_1,...,0,\mathbf{x}_N)~\stackrel{!}{=}~\varphi_0(\mathbf{x}_1,...,\mathbf{x}_N).
 \label{eq:multitimeinitialvalues}
\end{equation}
Alternatively, one can choose to prescribe initial data on a general set $\Sigma^N$ where $\Sigma$ is a space-like hypersurface.
One re-obtains a single-time wave equation
\begin{equation}
 i \frac{\partial \varphi}{\partial t}~=~\sum_{k = 1}^N H_k \, \varphi
 \label{eq:singletime}
\end{equation}
from eqs.\@ \eqref{eq:multitime} by using the chain rule together with eq.\@ \eqref{eq:connection}.\\
Because of the many simultaneous equations \eqref{eq:multitime}, questions of consistency arise. For the case that the $H_i$ can be regarded as self-adjoint operators on Hilbert space (then called ``partial Hamilonians''), it was shown in \cite{nogo_potentials} that the following consistency condition (first mentioned in \cite{bloch}) is necessary and sufficient for the existence of a unique solution on $\R^{4N}$ for the initial value problem \eqref{eq:multitimeinitialvalues}:
\begin{equation}
 \left[H_j - i \frac{\partial}{\partial t_j}, H_k - i \frac{\partial}{\partial t_k}\right] ~=~ 0~~\forall j,k.
\label{eq:kb}
\end{equation}
On the more natural domain $\mathscr{S}$, which has a non-empty boundary, boundary conditions may be required for the uniqueness of the solution. However, this depends sensitively both on the operators $H_i$ and on the space-time dimension \cite{svendsen,nogo_potentials,1d_model}. In the case that the $H_i$ cannot be regarded as operators on Hilbert space, e.g.\@ because they contain multiplication or differential operators with respect to time, different techniques have to be applied to prove the existence and uniqueness of a solution \cite{qftmultitime,1d_model}.

\subsection{Currents, probability density and Hilbert spaces} \label{sec:multitimecurrents}
In order for the multi-time equations to make contact to statistical predictions and to the formalism of non-relativistic QM, it is essential that \eqref{eq:multitime} implies the existence of a positive-definite tensor current $j^{\mu_1... \mu_N}_\psi(x_1,...,x_N)$ as a bilinear function of the wave function. The current is supposed to be conserved as a consequence of the wave equations, i.e.\@
\begin{equation}
 \partial_{k,\mu_k} j^{\mu_1...\mu_k... \mu_N}_\psi(x_1,...,x_N) ~=~0,~~k = 1,..., N,
 \label{eq:currentcons}
\end{equation}
 where $\partial_{k,\mu_k}$ is shorthand for $\frac{\partial}{\partial x_k^{\mu_k}}$.\\
In free multi-time $N$-particle Dirac theory, i.e.\@ for
\begin{equation}
 H_k \equiv H_k^{\rm Dirac} := -i \gamma_k^0 \gamma_k^j \partial_{k,j} + m_k \gamma_k^0
 \label{eq:freedirachamiltonians}
\end{equation}
one has:
\begin{equation}
 j^{\mu_1...\mu_N}_\psi(x_1,...,x_N) ~=~ \overline{\psi}(x_1,...,x_N) \gamma_1^{\mu_1} \cdots \gamma_N^{\mu_N}\psi(x_1,...,x_N),
 \label{eq:freediraccurrent} 
\end{equation}
where $\overline{\psi} = \psi^\D \gamma_1^0 \cdots \gamma_N^0$.\\
The positive component of $j_\psi$ then is $j_\psi^{0...0} = \psi^\D \psi$, i.e.\@ the usual $|\psi|^2$ density. Experimental predictions for an equal-time hypersurface $\Sigma_t$ are based on this density. More generally, the adequate probability density $\rho_\Sigma$ for a space-like hypersurface with normal covector field $n(x)$ is expected to be \cite{hbd}:
\begin{equation}
 \rho_\Sigma(x_1,...,x_N)~=~ j^{\mu_1...\mu_N}_\psi(x_1,...,x_N) \, n_{\mu_1}(x_1) \cdots n_{\mu_N}(x_N).
 \label{eq:generaldensity}
\end{equation}
We emphasize that the existence of the tensor current $j_\psi$ is the central requirement for a multi-time theory to make sense for physics. Even though it may seem only a minimal operational requirement, it is in fact sufficient to formulate a full realistic relativistic quantum theory \cite{grwf,rel_grw,hbd}.\\
With this picture in mind, one can define natural $N$-particle Hilbert spaces associated with a space-like hypersurface $\Sigma$ \cite{dirk_phd,hbd_subsystems}:
\begin{equation}
 \mathcal{H}_\Sigma^{(N)}~:=~ \left(L^2(\Sigma^N) \otimes \C^k,\,\langle \cdot, \cdot \rangle_\Sigma \right).
 \label{eq:hilbertspace}
\end{equation}
An appropriate scalar product $\langle \cdot, \cdot \rangle_\Sigma$ can be defined if there exists a bilinear form of two wave functions that generalizes the tensor current $j_\psi$, i.e.\@ if there exists
\begin{equation}
 j^{\mu_1... \mu_N}[\psi_1,\psi_2](x_1,...,x_N)~~~{\rm with}~~~\partial_{k,\mu_k} j^{\mu_1...\mu_k... \mu_N}[\psi_1,\psi_2](x_1,...,x_N)~=~0, ~~k = 1,..., N
\end{equation}
for all solutions $\psi_1,\psi_2$ of \eqref{eq:multitime}, and with
\begin{equation}
 j^{\mu_1... \mu_N}[\psi,\psi]~=~j_\psi^{\mu_1... \mu_N}.
 \label{eq:currentconnection}
\end{equation}
For free multi-time Dirac theory, we have:
\begin{equation}
 j^{\mu_1... \mu_N}[\psi_1,\psi_2](x_1,...,x_N) ~=~ \overline{\psi}_1(x_1,...,x_N) \gamma_1^{\mu_1} \cdots \gamma_N^{\mu_N} \psi_2(x_1,...,x_N).
 \label{eq:freediraccurrent2} 
\end{equation}
Then the natural choice\footnote{Note that the natural range of integration is $(\Sigma_1 \times \cdots \times \Sigma_N) \cap \mathscr{S}$, expressing that the particles are always within the domain. However, as discussed in \cite{1d_model}, for $d > 1$ space dimensions the difference to \eqref{eq:scalarprod} is only a zero-measure set.} for the scalar product is given by:
\begin{equation}
 \langle \phi, \chi \rangle_\Sigma~:=~ \int_{\Sigma^N} d \sigma(x_1) \wedge \cdots \wedge d \sigma(x_N) \, j^{\mu_1... \mu_N}[\phi,\chi](x_1,...,x_N) \, n_{\mu_1}(x_1) \cdots n_{\mu_N}(x_N).
 \label{eq:scalarprod}
\end{equation}
One can easily verify that for $j[\psi_1,\psi_2]$ given by \eqref{eq:freediraccurrent2} the scalar product reduces to the familiar expression $\langle \phi, \chi \rangle = \int d^3 x_1 \cdots d^3 x_N \, \phi^\D \chi$ for an equal-time hypersurface $\Sigma_t$.\\
Furthermore, the scalar product has physical meaning by its connection with $\rho_\Sigma$ from eq.\@ \eqref{eq:generaldensity}. This can be seen as follows: Define
\begin{equation}
 \| \psi\|_\Sigma~:= \sqrt{\langle \psi, \psi\rangle_\Sigma}.
\end{equation}
Let $A \subset \Sigma^N$ and let $\id_A$ denote the indicator function of the set $A$. If $\| \psi \|_\Sigma = 1$, then by eq.\@ \eqref{eq:generaldensity}, $\| \id_A \psi\|^2_\Sigma$ gives the probability for a spatio-temporal configuration of $N$ particles on $\Sigma$ to be in $A$ \cite{hbd_subsystems}.\\
We note that a similar expression for the scalar product has been suggested by Rizov, Sazdjian and Torodorov \cite{rizov_tensor_currents,sazdjian_scalarprod}. They define:
\begin{align}
 (\phi,\chi) ~:=~ \int_{\Sigma_1}  d \sigma(x_1)  \cdots \int_{\Sigma_N} d \sigma(x_N) \, j^{\mu_1... \mu_N}[\phi,\chi](x_1,...,x_N) \, n_{\mu_1}(x_1) \cdots n_{\mu_N}(x_N).
 \label{eq:scalarprod_sazdjian}
\end{align}
The crucial difference to eq.\@ \eqref{eq:scalarprod} is that Rizov et al. allow the space-like hypersurfaces $\Sigma_k$ to be different. This is problematic because according to the argument at the beginning of this section, a multi-time wave function is naturally only defined on the set $\mathscr{S}$ of space-like configurations. However, an element of $\Sigma_1 \times  \cdots \times \Sigma_N$ is in general not a space-like configuration.\\
Furthermore, there is no reason why then $\sqrt{(\psi,\psi)}$ should be equal to unity. This is because a crossing probability $\rho_\Sigma$ can in general only be defined on sets of the form $\Sigma^N$. The physical reason for this is that a time-like configuration may actually correspond to two points on the world-line of the same particle. Besides these physical arguments, the mathematical consistency of the multi-time equations \eqref{eq:kb} on the whole of $\R^{4N}$ may lead to very restrictive conditions on the possible operators $H_i$, e.g.\@ excluding a multi-time formulation of QED models which (disregarding UV divergencies) is possible on $\mathscr{S}$ \cite{bloch,tomonaga,qftmultitime}. The domain $\mathscr{S}$ also plays an important role for the possibility to rigorously introduce relativistic interactions by boundary conditions at the space-time points of coincidence in a simple $1+1$-dimensional model \cite{1d_model,nt_model}.\\
Returning to the Hilbert spaces $\mathcal{H}_\Sigma^{(N)}$, we can understand a multi-time system \eqref{eq:multitime} on $\mathscr{S}$ to define a unitary evolution between different space-like hypersurface $\Sigma,\Sigma'$ as follows \cite{qftmultitime,hbd_subsystems}: Denote by $\psi_{|_\Sigma}$ the restriction of a solution of \eqref{eq:multitime} to $\Sigma$. Then
\begin{equation}
 U_{\Sigma \rightarrow \Sigma'}: \mathcal{H}_\Sigma^{(N)} \rightarrow \mathcal{H}_{\Sigma'}^{(N)},~~~\psi_{|_\Sigma} \mapsto \psi_{|_{\Sigma'}}
 \label{eq:unitaryevolution}
\end{equation}
defines a map with $ \| \psi_{|_\Sigma} \|_\Sigma = \| U_{\Sigma \rightarrow \Sigma'} \psi_{|_{\Sigma'}} \|_{\Sigma'} = \| \psi_{|_{\Sigma'}} \|_{\Sigma'}$ \cite{hbd_subsystems}. This way of formulating the time evolution puts the multi-time system of equations in the foreground and yields the Hilbert space picture as a byproduct. Note that reversing the train of thought and first defining Hilbert spaces $\mathcal{H}_\Sigma^{(N)}$ and the map $U_{\Sigma \rightarrow \Sigma'}$ does not in general yield a multi-time wave function on $\mathscr{S}$ (which may be necessary for a consistent interpretation \cite{grwf,rel_grw,hbd,rel_bm}). This can be seen as follows: Pick two space-like hypersurfaces $\Sigma \neq \Sigma'$ with $\Sigma \cap \Sigma' \neq \emptyset$. Then the Hilbert spaces $\mathcal{H}_\Sigma^{(N)}$ and $\mathcal{H}_{\Sigma'}^{(N)}$ are essentially different and for $\phi \in \mathcal{H}_\Sigma^{(N)}$ we may have $\phi(q) \neq (U_{\Sigma \rightarrow \Sigma'} \phi)(q)$.\\
The connection to the Hilbert space setting for the Schrödinger equation is given by restricting to equal-time hypersurfaces $\Sigma_t$ in a distinguished frame and identifying all Hilbert spaces $\mathcal{H}_{\Sigma_t}^{(N)}$ for different $t$ \cite{qftmultitime}. This is possible for flat hypersurfaces without changing the scalar product. Also, because of this identification, the above-mentioned problem cannot occur because trivially there do not exist $t \neq t'$ with $\Sigma_t \cap \Sigma_{t'} \neq \emptyset$.

\subsection{A no-go theorem for potentials} \label{sec:nogotheorempotentials}
For classical mechanics, the ``no interaction theorem'' seems to rule out a sensible physical theory of $N$ directly interacting particles in the Hamiltonian framework \cite{nointeraction,nointeraction2}. Presupposing canonical quantization rules, this may also have significance for a relativistic quantum theory. There do, however, remain doubts if this connection is robust enough or if a quantum theory does not simply offer further possibilities (see \cite{multitime_classical_quantum} for a discussion). Considering the Dirac equation, one may especially think of spin which does not have a classical analog. In view of the severity of a no-go theorem, it is important that its assumptions exactly fit the appropriate situation. A no-go theorem appropriate for multi-time Dirac equations was recently proven in \cite{nogo_potentials}:\\
Let the operators $H_k$ in eq.\@ \eqref{eq:multitime} be of the form
\begin{equation}
 H_k~=~ H_k^{\rm Dirac} + W_k(x_1,...,x_N)
 \label{eq:hkpotentials}
\end{equation}
where $H_k^{\rm Dirac}$ is defined in eq.\@ \eqref{eq:freedirachamiltonians} and $W_k$ is a matrix-valued function of its arguments which only acts on the spin index of the $k$-th particle\footnote{This assumption is needed for the free Dirac current \eqref{eq:freediraccurrent} to be conserved in the presence of the potentials.}. Then the consistency conditions \eqref{eq:kb} are satisfied if and only if the multi-time equations are gauge-equivalent to the case $W_k = W_k(x_k)$, i.e.\@ of purely external potentials.\\
It is therefore an interesting challenge to find different mechanisms for relativistic interactions. Multi-time QFT models provide further possibilities for interaction terms by using creation and annihilation operators \cite{dfp,tomonaga,schwinger,qftmultitime} but they encounter the problem of UV divergencies. For $1+1$ dimensions, one can construct a rigorous, interacting model by boundary conditions on the set of coincidence points of two particles in space-time \cite{1d_model,nt_model}. The 2BD equations offer a third possibility by allowing the total momentum operator in the ``potential'' terms. This departs from the form \eqref{eq:multitime} of the multi-time equations in that time derivatives are present in the interaction terms. Consequently, the system is not ``Hamiltonian'' and thus the ``no interaction theorem'' cannot be applied. The option of the 2BD equations seems especially attractive because the construction works for $1+3$ dimensions and does not lead to any known 
 UV divergencies. This constitutes 
a strong motivation to analyze if these equations can indeed be used as the basis of a self-contained (and in this sense fundamental) relativistic multi-time theory for two particles.

\section{The Two-Body Dirac equations as multi-time evolution equations}

In this section, we introduce the idea underlying the 2BD equations. Starting from a general form of the multi-time equations which admits differential operators of any order in the interaction terms, we analyze the implications of a necessary compatibility condition when combined with Poincaré invariance (sec.\@ \ref{sec:implicationskb}). The central features are that the interaction terms are related in a specific way and that a certain variable $x_\perp$ appears which generalizes the spatial distance in the center of momentum frame in a subtle and covariant way.\\
For definiteness, we then specialize to a general class of the 2BD equations for fermion-antifermion systems which was first presented by Sazdjian in \cite{sazdjian_2bd} (sec.\@ \ref{sec:2bdfermionantifermion}). The focus is on the case of Dirac (spin-$\tfrac{1}{2}$) particles because for Klein-Gordon particles already in the one-particle case the candidate probability density can be negative \cite{schweber}. The choice of the particular class of 2BD equations is made because (a) its compatibility can be seen straightforwardly, (b) it can in some sense be derived from the Bethe-Salpeter equation \cite{sazdjian_bs} and (c) there exists an analysis of its tensor currents \cite{sazdjian_scalarprod}. We emphasize that the focus on the this class of the 2BD equations is not very restrictive as there exists a formal relation of Sazdjian's 2BD equations and those proposed by Crater and Van Alstine \cite{crater_wf,mourad_sazdjian}. The main features of these two classes of 2BD equations 
are shared, such as a dependence of the potential terms on the total momentum operator.\\
Having introduced the equations, we comment on the question of how one can understand them as evolution equations for a multi-time wave function (sec.\@ \ref{sec:preliminaryunderstanding}). As these equations leave the standard framework for the existence and uniqueness theory of quantum mechanical wave equations in several respects, it is required to introduce at least a preliminary mathematical understanding.

\subsection{Notation}
We use the following abbreviations:
\begin{align}
 &x = x_1 - x_2,~~~X = (x_1 + x_2)/2, \nonumber\\
 &\hat{p}_{k,\mu} = i \frac{\partial}{\partial x_k^\mu},~~k=1,2,\nonumber\\
 &\hat{p} = (\hat{p}_1 - \hat{p}_2)/2,~~~\hat{P} = \hat{p}_1 + \hat{p}_2.
  \label{eq:notation}
\end{align}

\subsection{Implications of the consistency condition for the general form of the equations} \label{sec:implicationskb}

Consider the following general form of multi-time wave equations:
\begin{equation}
 D_i \psi(x_1,x_2)~=~0,~~i = 1,2.
\label{eq:generalform2bd}
\end{equation}
The crucial point is that in contrast to the form in eq.\@ \eqref{eq:multitime} (which led to the no-go theorem \cite{nogo_potentials}) we now allow the operators $D_i$ to be differential operators of \textit{any order}, including infinity.\\
Recall from sec.\@ \ref{sec:multitimederiv} that certain compatibility conditions have to be satisfied in order for the wave equations \eqref{eq:generalform2bd} to have solutions. However, because the $D_i$ are not necessarily first-order differential operators, eqs.\@ \eqref{eq:generalform2bd} cannot be cast into the form \eqref{eq:multitime}. Therefore, the consistency condition \eqref{eq:kb} is not appropriate.
For general operators $D_i$, one cannot expect to find a simple replacement of eq.\@ \eqref{eq:kb} which is also both necessary and sufficient. However, a necessary condition reads
\begin{equation}
 [D_1,D_2] ~=~ \lambda_1 D_1 + \lambda_2 D_2
 \label{eq:weak_kb}
\end{equation}
where at least formally the $\lambda_i$ can be operators.\\
To see that \eqref{eq:weak_kb} is a necessary condition, consider a solution $\psi$ of \eqref{eq:generalform2bd}. Assume that one cannot write $[D_1,D_2]$ in the form \eqref{eq:weak_kb}. In general, we then have $[D_1,D_2] = \lambda_1 D_1 + \lambda_2 D_2 + R$ where $R \psi \neq 0$. Thus: $[D_1,D_2] \psi = R \psi \neq 0$ in contradiction with the fact that $[D_1,D_2]\psi = 0$ trivially holds for any solution of \eqref{eq:2bdsazdjian} on which the action of $D_1 D_2$ and $D_2 D_1$ is well-defined.
\begin{remark}
 Note that in the case that the operators $D_i$ are of first order, condition \eqref{eq:weak_kb} bears some similarity with the consistency condition of Frobenius' theorem from differential topology. However, it was discussed in \cite[sec.\@ 2.4]{nogo_potentials} that the conditions for multi-time wave functions are, even for first-order operators, different from the statement of Frobenius' theorem, one of the reasons being the number of components of $\psi$.\\
 Furthermore, note that \eqref{eq:weak_kb} applied to the first-order multi-time equations \eqref{eq:multitime} seems to lead to a weaker condition than the previous condition \eqref{eq:kb}. For \eqref{eq:weak_kb} it is sufficient for the right hand side to be a linear combination of the operators $\left(i \tfrac{\partial}{\partial t_k} - H_k \right)$ instead of having to vanish.
 At first glance, the result in \cite{nogo_potentials}, according to which \eqref{eq:kb} is a necessary and sufficient condition thus seems to deem \eqref{eq:weak_kb} a too weak condition. However, \cite{nogo_potentials} is tied to the case that the operators $H_k$ are operators on Hilbert space (or more generally operator-valued functions of the time variables). As the operators $D_i$ contain time derivatives and therefore cannot be regarded as operators on Hilbert space, it does not follow from \cite{nogo_potentials} that the right hand side of eq.\@ \eqref{eq:weak_kb} has to vanish. Still, the question of a necessary \textit{and} sufficient condition for the system of equations \eqref{eq:generalform2bd} remains open.
\end{remark}
The important question now is: \textit{Which operators $D_i$ satisfy condition \eqref{eq:weak_kb}?} In order to reduce this question to a tractable form, we assume with \cite{sazdjian_2bd} that operators $D_i$ take the following form:
\begin{equation}
 D_i ~=~ D_{i,0} + D_{3-i,0} \hat{V},~~i = 1,2,
\label{eq:formdi}
\end{equation}
where $D_{i,0}$ are the operators of the free equation and $\hat{V}$ is an operator the structure of which is yet to be determined.  Generally, it may depend on $x$, $\hat{p}$ and $\hat{P}$ as well as the gamma matrices $\gamma_1^\mu, \gamma_2^\nu$. A dependence on $X$ is excluded if one aims at a Poincaré invariant theory. Note that \eqref{eq:formdi} introduces a relation between the interaction terms in the two wave equations \eqref{eq:generalform2bd}. We remark that the form \eqref{eq:formdi} may imply that the wave function of eq.\@ \eqref{eq:generalform2bd} cannot be directly identified with the wave function of, say, the Breit equation, or the class of 2BD equations of Crater and Van Alstine. A wave function transformation may be necessary to relate the two types of wave functions (see \cite{2bd_hyperbolic,mourad_sazdjian}). Notwithstanding, we analyze the theory resulting from \eqref{eq:formdi} on its own terms.\\
Using the form \eqref{eq:formdi}, we obtain:
\begin{align}
 [D_1,D_2]~&=~ D_{1,0} D_{2,0} + D_{1,0}^2 \hat{V} + D_{2,0} \hat{V} D_{2,0} + D_{2,0} \hat{V} D_{1,0} \hat{V}\nonumber\\
 &~~~~-D_{2,0} D_{1,0} - D_{2,0}^2 \hat{V} - D_{1,0} \hat{V} D_{1,0} - D_{1,0} \hat{V} D_{2,0} \hat{V}.
 \label{eq:commutatorcalc1}
\end{align}
As the $D_{i,0}$ are supposed to correspond to the operators in a free wave equation (acting only on the coordinates and spin indices of the $i$-th particle), we have $[D_{1,0},D_{2,0}] = 0$ and the first summands in the lines of eq.\@ \eqref{eq:commutatorcalc1} cancel.\\
Aiming to bring \eqref{eq:commutatorcalc1} into the form of the right hand side of eq.\@ \eqref{eq:weak_kb}, we calculate the expression
\begin{align}
 -[D_{1,0},\hat{V}] D_1 + [D_{2,0}, V] D_2~&=~ -D_{1,0} \hat{V} D_{1,0} - D_{1,0} \hat{V} D_{2,0} \hat{V} + \hat{V} D_{1,0}^2\nonumber\\
 &~~~~~+ D_{2,0} \hat{V} D_{2,0} + D_{2,0} \hat{V} D_{1,0} \hat{V} - \hat{V} D_{2,0}^2.
 \label{eq:commutatorcalc2}
\end{align}
Comparing eqs.\@ \eqref{eq:commutatorcalc1} and \eqref{eq:commutatorcalc2}, condition \eqref{eq:weak_kb} is satisfied if
\begin{equation}
 [D_{1,0}^2-D_{2,0}^2,\hat{V}] ~=~ 0.
 \label{eq:dicondition}
\end{equation}
Specializing to the Dirac case, we evaluate \eqref{eq:dicondition} for
\begin{equation}
 D_{i,0} ~=~ \gamma_i \cdot \hat{p}_i - m_i.
 \label{eq:diraccase}
\end{equation}
Then: $D_{i,0}^2 = \hat{p}_i^2 + m_i^2$ and eq.\@ \eqref{eq:dicondition} reduces to:
\begin{equation}
 [\hat{p}_1^2-\hat{p}_2^2,\hat{V}]~=~0.
\end{equation}
 Rewriting this equation using total and relative momentum operators yields:
 \begin{equation}
  [\hat{P} \cdot \hat{p},\hat{V}]~=~0.
 \end{equation}
 Now, because of Poincaré invariance, $\hat{V}$ must not depend on $X$. Thus, we arrive at the condition
 \begin{equation}
  P^\mu \frac{\partial \hat{V}}{\partial x^\mu}~\stackrel{!}{=}~0.
 \label{eq:xperpcond}
 \end{equation}
 To further evaluate eq.\@ \eqref{eq:xperpcond}, note that the only Poincaré-invariant quantities involving $x$ are of the form $x \cdot q$ where $q$ is a quantity transforming as a $4$-vector (e.g.\@ $x, \hat{P}, \hat{p}, \gamma_1,\gamma_2$). Thus, $\frac{\partial \hat{V}}{\partial x^\mu} \propto q_\mu$ and \eqref{eq:xperpcond} requires $q \perp \hat{P}$ in the Minkowski sense. This can in general only be achieved if $q_\mu$ has the form
 \begin{equation}
  q_\mu ~=~\hat{\pi}^\nu_\mu\tilde{q}_\nu
 \label{eq:projectorcond}
 \end{equation}
 where $\hat{\pi}^\nu_\mu := \left( 1 - \frac{\hat{P}_\mu \hat{P}^\nu}{\hat{P} \cdot \hat{P}}\right)$ is (an operator version of) the projection operator on the subspace ``orthogonal to $\hat{P}$'' and $\tilde{q}$ again transforms as a vector.\\
 The most important consequence of this can be seen by considering the case that $\tilde{q} = x$. Then eq.\@ \eqref{eq:projectorcond} implies that the only dependence of $\hat{V}$ on $x$ may be via
 \begin{equation}
  \hat{x}_\perp^\mu ~:=~ \hat{\pi}_{\nu}^\mu x^\nu
  \label{eq:xperpdef}
 \end{equation}
 To see this in detail, note that $x \cdot q = x \cdot \hat{x}_\perp = x \cdot (\hat{\pi} x) = x \cdot (\hat{\pi}^2 x) = \hat{x}_\perp \cdot \hat{x}_\perp$.

\paragraph{On the meaning of $x_\perp$:} In classical mechanics, the analog $x_\perp$ of the operator $\hat{x}_\perp$ (i.e.\@ where $\hat{P}$ in eq.\@ \eqref{eq:notation} is replaced by a time-like 4-vector $P$) has the following meaning. Consider the relative spatial coordinate $\mathbf{x} = \mathbf{x}_1 - \mathbf{x}_2$ between the two particles in the center of momentum (c.m.\@) frame, i.e.\@ the frame where the total momentum 4-vector takes the form $(P^0,0,0,0)$. Then $x_\perp = (0,\mathbf{x})$. In this way, one can see that $x_\perp$ is the covariantization of $(0,\mathbf{x})$.\\
Note that generalizing a non-relativistic law by the replacement $(0,\mathbf{x}) \rightarrow x_\perp$ would seem suspicious. If $P$ were the total momentum of the total system, it would be inacceptable because in this context total quantities do not have any physical meaning but are only used to define coordinate systems. However, the use of $x_\perp$ is restricted to autonomous two-particle systems\footnote{See \cite{hbd_subsystems} for an analysis of when an autonomous subsystem description is possible for relativistic systems with spin.} which are to be thought of as subsystems of a larger system. Then, the total momentum $P$ is meaningful. Yet, one might object against the use of any preferred frame, even if it is dynamically preferred, such as the c.m.\@ frame here. However, this criticism is alleviated by the fact that the replacement $(0,\mathbf{x}) \rightarrow x_\perp$ is never used in the derivation of the necessity of $x_\perp$. Rather, the crucial starting point is the for
 m \eqref{eq:formdi} of 
the 
operators $D_i$ -- which is far from directly assuming a covariantization of a non-relativistic law of motion\footnote{A related subtlety of the notion of Lorentz invariance was critically discussed by Bell in \cite{bell_qftbeables}.}.

\paragraph{Further remarks:}
\begin{enumerate}
 \item One can also motivate the necessity of $\hat{x}_\perp$ very concisely in the context of two-body Klein-Gordon equations of the form $(\hat{p}_i^2 - m_i^2 - \hat{V})\psi = 0,~i = 1,2$ where $\hat{V}$ is a scalar and Poincaré invariant potential (operator) \cite[sec.\@ II]{2bd_scalar_vector}. However, this derivation is of limited significance for the approach to the 2BD equations taken here because the square of the operators \eqref{eq:formdi} does not in general yield $\hat{p}_i^2 - m_i^2 - \hat{V}$ with a scalar potential $\hat{V}$\footnote{Note that starting from a different point, Crater and Van Alstine were in fact able to derive 2BD equations with scalar interactions as ``square roots'' of corresponding scalar interacting two-body Klein-Gordon equations \cite{2bd}.}. 
 \item The necessity of $\hat{x}_\perp$ does not follow from the connection of the 2BD equations with the Bethe-Salpeter equation (see \cite{sazdjian_bs}). Rather, the insight that $\hat{x}_\perp$ is necessary to formulate differential equations of the type \eqref{eq:generalform2bd} is itself used to make the so-called ``relativistic instantaneous approximation'' which creates a manifest $\hat{x}_\perp$-dependence of the potential terms.
\end{enumerate}

\subsection{Two-Body Dirac equations for fermion-antifermion systems} \label{sec:2bdfermionantifermion}

For the rest of the paper, we now specialize to an important class of 2BD equations which is almost identical to one discussed above: the case of spin-$\tfrac{1}{2}$ particle-antiparticle pairs. This case is particularly relevant as one aims at a theoretical description of mesons and their spectra \cite{crater_applications,crater_quark_model,jallouli_sazdjian_98,crater_meson_spectroscopy}. Equations for fermions with the same charge are, on the other hand, not believed to describe bound states (as are the equations below) and therefore not to lead to particularly interesting subsystem dynamics.\\
The class of 2BD equations for particle-antiparticle pairs (first introduced by Sazdjian in \cite[sec.\@ VI]{sazdjian_2bd}) is given by
\begin{align}
 D_1 \psi(x_1, x_2) ~&\equiv~ \left[ \gamma_1 \cdot \hat{p}_1 - m_1 -(- \gamma_2 \cdot \hat{p}_2 + m_2)\hat{V}\right] \psi(x_1,x_2) ~=~ 0,\nonumber\\
 D_2 \psi(x_1, x_2) ~&\equiv~ \left[ \gamma_2 \cdot \hat{p}_2 + m_2 +(\gamma_1 \cdot \hat{p}_1 + m_1)\hat{V}\right] \psi(x_1,x_2) ~=~ 0.
 \label{eq:2bdsazdjian}
\end{align}
Here, $\psi$ is a $16$-component wave function. According to standard sign conventions, particle 2 is the anti-particle.\\
The form of the equations is motivated similarly to the approach via eqs.\@ \eqref{eq:generalform2bd} and \eqref{eq:formdi}, the only difference being that one has to account for the symmetries of the fermion-antifermion system. More precisely, the first of the equations \eqref{eq:2bdsazdjian} has to be obtained from the second via charge conjugation and mass exchange \cite[p. 3411]{sazdjian_2bd}. This changes some signs when compared with the form of the operators $D_i$ in eq.\@ \eqref{eq:formdi}.\\
$\hat{V}$ is an operator which may depend on $\hat{P}, \hat{p}, \hat{x}_\perp$ and the $\gamma$-matrices in a Poincaré invariant manner. The symmetry of the fermion-antifermion system demands \cite[p. 3411]{sazdjian_2bd}:
\begin{equation}
 \hat{V}(1,2;\gamma_1,\gamma_2) ~=~ \hat{V}(2,1;-\gamma_2,-\gamma_1)
\label{eq:vcond1}
\end{equation}
where ``$1 \leftrightarrow 2$'' indicates the exchange of particle labels in quantities like $\hat{p}_i, x_i$ (but not the $\gamma$-matrices). We remark that here and in the following the notation $\hat{V}(...)$ is only meant to emphasize possible dependencies on certain variables. $\hat{V}$ is always the same operator.\\
Then the 2BD equations \eqref{eq:2bdsazdjian} satisfy also the compatibility condition \eqref{eq:weak_kb} because the following relation holds \cite[p. 3412]{sazdjian_2bd}:
\begin{equation}
 [D_1,D_2] ~=~ -[\gamma_1\cdot \hat{p}_1,\hat{V}] D_1 + [\gamma_2\cdot \hat{p}_2,\hat{V}] D_2.
 \label{eq:weak_kb_sazdjian}
\end{equation}
Special choices of $\hat{V}$ may yield $[D_1,D_2] = 0$ \cite[sec.\@ VII]{sazdjian_2bd}.

\subsection{Basic mathematical questions} \label{sec:basicquestions}
Due to the dependence of $\hat{V}$ on $\hat{P}$, the 2BD equations \eqref{eq:2bdsazdjian} are of infinite order both in space and time coordinates. This immediately raises difficult mathematical questions which are generally not addressed in the literature (compare e.g.\@ \cite{sazdjian_2bd,2bd}). Besides the question of the compatibility of the 2BD equations which was already discussed in sec.\@ \ref{sec:implicationskb}, one may ask:
\begin{enumerate}
 \item What are appropriate initial data?
 \item What is an adequate space of solutions?
\end{enumerate}
Concerning 1., note that in contrast to a first-order multi-time system \eqref{eq:multitime}, one does not expect initial data to consist only of prescribing the wave function for configurations on a space-like hypersurface like in \eqref{eq:multitimeinitialvalues}. For wave equations of $n$-th order one would rather expect that also $(n-1)$-th time derivatives have to be prescribed. If this analogy extended to infinite order, this understanding of time evolution for the 2BD equations could not make sense since prescribing all derivatives of an (analytic) function on a Cauchy surface is equivalent to writing down the solution on the whole of $\R^{8}$ (for two particles). One may, however, hope that the fact that the infinite order only arises from the dependence of $\hat{V}$ on the total momentum operator may help to identify sensible initial data (see section \ref{sec:preliminaryunderstanding}).\\
With respect to point 2, note that the possible spaces of initial data are usually a good starting point for defining Hilbert spaces on which at least the non-relativistic existence and uniqueness theory is usually based. Due to the occurrence of powers of $\hat{P}$ to infinite order, and therefore of time derivatives, it is clear that this setting cannot be used without mayor changes. Moreover, the considerations in section \ref{sec:multitimecurrents} show that the natural Hilbert spaces depend via the scalar product \eqref{eq:scalarprod} (and corresponding statements about self-adjointness etc.) on the form of the conserved tensor current of the theory \eqref{eq:currentcons}. The question of conserved currents for the 2BD theory will be addressed in section \ref{sec:currents}.

\subsection{A preliminary mathematical understanding} \label{sec:preliminaryunderstanding}
In this subsection, we propose a way how one can understand the 2BD equations in a preliminary way for superpositions of eigenfunctions of the total momentum operator.\\
Assume that the only momentum dependence of $\hat{V}$ is via $\hat{P}$ (explicitly via $P^2$ or implicitly via $\hat{x}_\perp$). We write: $\hat{V} = \hat{V}(\hat{x}_\perp,\hat{P})$. Let $\psi_P$ be an eigenfunction of $\hat{P}$, i.e.\@ a function of the form
\begin{equation}
 \psi_P(x_1,x_2)~=~ \tilde{\psi}(x) e^{-i P \cdot X}.
 \label{eq:totalmomentumeigenfunction}
\end{equation}
Then
\begin{equation}
 \hat{V}(\hat{x}_\perp,\hat{P}) \psi_P ~\equiv~ V(x_\perp,P) \psi_P,
 \label{eq:vmultiplication}
\end{equation}
where $V(x_\perp,P)$ is the matrix-valued function which is obtained by replacing $\hat{P}$ in $\hat{x}_\perp$ with its eigenvalue $P$. In this way, we can regard $\hat{V}$ as a multiplication operator.\\
For $\psi_P$, eqs.\@ \eqref{eq:2bdsazdjian} then constitute a first order system of differential equations, analogous to \eqref{eq:multitime}. The analogy with \eqref{eq:multitimeinitialvalues} then suggests that adequate initial data are of the form of prescribing $\psi_P(x_1,x_2)$ on a space-like hypersurface, e.g.\@ for $x_1^0 = x_2^0 = 0$:
\begin{equation}
 \psi_P(0,\mathbf{x}_1,0,\mathbf{x}_2)~\stackrel{!}{=}~\tilde{\psi}_0(\mathbf{x}) e^{i \mathbf{P} \cdot \mathbf{X}},~~\mathbf{x} \in \R^3.
 \label{eq:eigenfunctioninitialdata}
\end{equation}
The role of the two eqs.\@ \eqref{eq:2bdsazdjian} then is to (a) time-evolve $\tilde{\psi}$ in $x^0$ and (b) determine $P^0$.\\
More generally, one should consider superpositions of eigenfunctions of $\hat{P}$. These functions are necessary to describe localized wave packets\footnote{Note that the issue of localization is not as problematic for the 2BD equations as e.g.\@ in relativistic quantum field theory where the Hamiltonian is assumed to be bounded from below. The reason is that, as with the single-particle Dirac equation, for the 2BD equations negative eigenvalues of the energy-momentum operators are possible.} on the configuration space of two particles. Let $Z$ denote further quantities, e.g.\@ the relative momentum eigenvalues, which classify a suitable space of ``relative coordinate wave functions'' $\tilde{\psi}(x)$. Then:
\begin{equation}
 \psi(x_1,x_2)~=~ \int dZ \int d^3 \mathbf{P} \, c(\mathbf{P},Z) \tilde{\psi}_Z(x) e^{-i P \cdot X},
 \label{eq:momentumeigenfunctionsuperposition}
\end{equation}
where it is understood that each $P^0$ is determined by eqs.\@ \eqref{eq:2bdsazdjian} by demanding that $\tilde{\psi}_Z(x) e^{-i P \cdot X}$ be a solution for every $Z, \mathbf{P}$.\\
The further strategy of the paper is the following: Setting aside all further mathematical questions\footnote{A good starting for the question of existence and uniqueness might be to first specialize on the case of a total momentum eigenfunction \eqref{eq:totalmomentumeigenfunction}. One may then hope that via \eqref{eq:momentumeigenfunctionsuperposition} a suitable space of solutions can be constructed.}, we assume that solutions of the form \eqref{eq:momentumeigenfunctionsuperposition} exist, at least for superpositions of finitely many eigenvalues of $\hat{P}$. This permits us to analyze the central physical question if there exist adequate conserved currents for the 2BD equations.

\section{The question of current conservation} \label{sec:currents}
Recall the central place of the tensor current $j^{\mu \nu}[\psi_1,\psi_2]$ and especially $j^{\mu \nu}[\psi,\psi]$ in the general structure of a multi-time theory (sec.\@ \ref{sec:multitimecurrents}). The importance of $j$ has also been recognized by various authors in the context of the Two-Body Dirac equations, in particular for the question how to construct scalar products and corresponding Hilbert spaces, see \cite{rizov_tensor_currents,longhi_lusanna_86} for the spin-less Klein-Gordon case and \cite{sazdjian_2bd,sazdjian_scalarprod} for the Dirac case with spin.\\
Here, we first review previous results for the 2BD equations, adding details and clearly stating critical assumptions (sec.\@ \ref{sec:previousresults}). It turns out that the free Dirac current is not conserved and that one has to construct suitable replacements. While replacements can be found, they are neither unique nor simple. We follow Sazdjian \cite{sazdjian_scalarprod} to pick a particular one in order to be able to further analyze the resulting theory. The expression for the scalar product (see eq.\@ \eqref{eq:scalarprod}) is in general not positive definite and a comprehensive analysis of the associated problems has (to our best knowledge) not been performed yet. Therefore, we discuss whether the ``scalar product'' can be rendered positive definite by (a) restrictions on the function space (sec.\@ \ref{sec:restrictedfunctions}) and (b) restrictions on the potential terms (sec.\@ \ref{sec:restrictedpotentials}). The further implications of the respective changes are illustra
 ted in 
both cases. 

\subsection{Previous results} \label{sec:previousresults}

In the following we assume that $\hat{V}$ satisfies the following hermiticity condition:
\begin{equation}
 V^\D(x_\perp,P) ~=~ \gamma_1^0 \gamma_2^0 V(x_\perp,P) \gamma_1^0 \gamma_2^0
 \label{eq:hermiticitycond}
\end{equation}
where $V$ (without the hat) was introduced in eq.\@ \eqref{eq:vmultiplication}.\\

\noindent \textbf{Claim 1:} \textit{Let $\psi_P, \psi_{P'}$ be eigenfunctions of $\hat{P}$. Then the free Dirac tensor current $j_{\rm free}^{\mu \nu}[\psi_P,\psi_{P'}] = \overline{\psi}_P \gamma_1^\mu \gamma_2^\nu \psi_{P'}$ is conserved if and only if no interaction terms $-(- \gamma_2 \cdot \hat{p}_2 + m_2)\hat{V}$ and $(\gamma_1 \cdot \hat{p}_1 + m_1)\hat{V}$, respectively, are present in the 2BD equations.}

\begin{proof}
 Consider
\begin{equation}
 i \partial_{1,\mu} \left(\overline{\psi}_P \gamma_1^\mu \gamma_2^\nu \psi_{P'}\right) ~=~ -(\gamma_1^\mu \hat{p}_{1,\mu}\psi_P)^\D \gamma_1^0 \gamma_2^0 \gamma_2^\nu \psi_{P'} ~+~ \overline{\psi}_P  \gamma_2^\nu (\gamma_1^\mu  \hat{p}_{1,\mu}\psi_{P'}).
 \label{eq:freecurrentcons1}
\end{equation}
Denote\footnote{This replaces the previous notation $V(x_\perp,P)$ to fit in the equations.} $\hat{V} \psi_P$ by $V_P\psi_P$ -- which still contains a $x_\perp$-dependence. The first of the 2BD equations \eqref{eq:2bdsazdjian} yields:
\begin{equation}
 \gamma_1 \cdot \hat{p}_1 \, \psi_{P'}~=~ \left[ m_1 + (-\gamma_2\cdot \hat{p}_2 + m_2) V_{P'}\right] \psi_{P'}.
\label{eq:freecurrentcons2}
\end{equation}
Using the relations \eqref{eq:hermiticitycond} as well as $(\gamma_k^\mu)^\D = \gamma_k^0 \gamma_k^\mu \gamma_k^0$ it follows that
\begin{equation}
  (\gamma_1 \cdot \hat{p}_1 \, \psi_P)^\D ~=~ \overline{\psi}_P \left[  V_P(m_2 + \gamma_2\cdot \stackrel{\leftarrow}{\hat{p}_2}) + m_1\right] \gamma_1^0 \gamma_2^0
 \label{eq:freecurrentcons3}
 \end{equation}
where the arrow indicates the direction in which the derivative acts. Combining eqs.\@ \eqref{eq:freecurrentcons1}, \eqref{eq:freecurrentcons2} and \eqref{eq:freecurrentcons3}, we obtain:
\begin{equation}
  i \partial_{1,\mu} (\overline{\psi}_P \gamma_1^\mu \gamma_2^\nu \psi_{P'}) ~=~ \overline{\psi}_P \left[ -  V_P(m_2 + \gamma_2\cdot \stackrel{\leftarrow}{\hat{p}_2}) \gamma_2^\nu + \gamma_2^\nu (-\gamma_2\cdot \hat{p}_2 + m_2) V_{P'}  \right] \psi_{P'}.
 \label{eq:freecurrentcons4}
\end{equation}
We note the following points:
\begin{enumerate}
 \item The term with $m_1$ has dropped out.
 \item The term with $m_2$, i.e.\@ $-V_Pm_2 \gamma_2^\nu + \gamma_2^\nu m_2 V_{P'}$ yields zero only in the case that $V$ does not contain $\gamma_2$-matrices and for $P = P'$.
 \item Even in the latter case, the remaining term $-V_P\gamma_2\cdot \stackrel{\leftarrow}{\hat{p}_2} \gamma_2^\nu - \gamma_2^\nu \gamma_2\cdot \hat{p}_2 V_{P'}$ does not vanish because $[\gamma_2^\nu,\gamma_2 \cdot \hat{p}_2] \neq 0$.
 \item If $P \neq P'$ and $V$ is not constant, not even special choices for $V$ make the rhs. of \eqref{eq:freecurrentcons4} vanish. The appearance of both $V_P$ and $V_{P'}$ is unavoidable because the basic mechanism which allows the 2BD equations to circumvent the no-go theorems is the use of these momentum-dependent terms.
 \item The only case in which the free Dirac current is conserved is that the 2BD equations do not contain the interaction terms from the very beginning.
\end{enumerate}
Analogous reasoning for $-i \partial_{2,\nu} \left(\overline{\psi}_P \gamma_1^\mu \gamma_2^\nu \psi_{P'}\right)$ establishes the claim. \qed
\end{proof}

\noindent \textbf{Claim 2} (see \cite[p.1625]{sazdjian_scalarprod}): \textit{There exist currents $j_{\rm int}^{\mu \nu}[\psi_P,\psi_{P'}]$ which are conserved by the 2BD equations.}

\begin{proof}
 One looks for a term $j_{\rm add}^{\mu \nu}[\psi_P,\psi_{P'}]$ such that
 \begin{equation}
  j^{\mu \nu}_{\rm int}[\psi_P,\psi_{P'}]~:=~ j_{\rm free}^{\mu \nu}[\psi_P,\psi_{P'}] + j^{\mu \nu}_{\rm add}[\psi_P,\psi_{P'}]
  \label{eq:jint}
 \end{equation}
 is conserved. We leave away the square brackets $[\psi_P,\psi_{P'}]$ in the following for notational ease but emphasize that in order to be able to treat $V$ as a matrix this understanding is crucial. Defining
 \begin{equation}
  F_1^\nu := \partial_{1,\mu} j_{\rm free}^{\mu \nu},~~~F_2^\mu := \tfrac{\partial}{\partial x_2^\nu} j_{\rm free}^{\mu \nu},~~~F := \partial_{1,\mu} \partial_{2,\nu} j_{\rm free}^{\mu \nu}
  \label{eq:f}
 \end{equation}
 we see that $j_{\rm add}^{\mu \nu}$ has to be a solution of the equations
 \begin{equation}
  \partial_{1,\mu} j_{\rm add}^{\mu \nu} = - F_1^\nu,~~~\partial_{2,\nu} j_{\rm add}^{\mu \nu} = - F_2^\mu.
  \label{eq:jadddef}
 \end{equation}
 Such a solution is easy to find \cite[eq.\@ (3.19)]{sazdjian_scalarprod}. Let $G(x-x')$ be a Green's functions of the four-dimensional wave equation, i.e.\@
 \begin{equation}
   \square_x G(x-x')~=~ \delta^{(4)}(x-x').
 \end{equation}
 Then for any pair of Green's functions $G_i,~i = 1,2$, a solution of \eqref{eq:jadddef} is given by:
 \begin{align}
  j^{\mu \nu}_{\rm add}(x_1,x_2) ~:&=~-\partial_1^\mu \int d^4 x_1' \, G_1(x_1-x_1') F_1^\nu(x_1',x_2) -  \partial_2^\nu \int d^4 x_2' \, G_2(x_2-x_2') F_2^\mu(x_1,x_2') \nonumber\\
  &~~~~~ +  \partial_1^\mu  \partial_2^\nu \int d^4 x_1' d^4 x_2' \, G_1(x_1-x_1') G_2(x_2-x_2') F(x_1,x_2').
  \label{eq:jaddsolution} \qed
 \end{align}
\end{proof}

\begin{remark}
 \begin{enumerate}
  \item From eq.\@ \eqref{eq:jaddsolution} it is obvious that $j_{\rm add}^{\mu \nu}$ \textit{is not defined uniquely.} One has to make a choice of the Green's functions $G_i,~i = 1,2$. Sazdjian's choice is\footnote{Sazdjian only uses eq.\@ \eqref{eq:jaddsolution} with $G_1 \equiv G_2$.} $G_i \equiv G_A,~i = 1,2$ where $G_A$ is the advanced Green's function, with the reason that this would be ``the only solution of \eqref{eq:jadddef} which vanishes when the interaction is switched off'' \cite[p. 1625]{sazdjian_scalarprod}.
  \item The construction of $j_{\rm int}$ is very general (one might even say too general) since it would have worked for any $F_1^\nu, F_2^\mu$ defined by eq.\@ \eqref{eq:f}. So what is the significance of the existence of conserved $j_{\rm int}$'s? A good answer would be to point out, for example, a unique current with the required properties like a positive component (see sec.\@ \ref{sec:multitimecurrents}). However, a \textit{general argument} why any of the possible definitions $j_{\rm int}$ should yield a positive definite current simply does not exist.
  \item Nevertheless, one may ask the question whether given \textit{further assumptions} the currents are positive definite. Further assumptions might even be plausible, for example if they concern special potentials $\hat{V}$. In the end, it is only important that the currents are positive definite for realistic choices of $\hat{V}$. An approach involving further assumptions was chosen by Sazdjian \cite{sazdjian_scalarprod} which we shall critically review next.
 \end{enumerate}
\end{remark}
Sazdjian's paper does not directly deal with the question whether the currents $j_{\rm int}$ are positive definite but rather with the one whether the associated scalar product (see eqs.\@ \eqref{eq:scalarprod}, \eqref{eq:scalarprod_sazdjian}) is positive definite. However, these two questions are equivalent as long as the wave functions admitted in the construction of the tensor currents (and in the scalar product) are not subject to restrictions which forbid localized wave packets.\\
Sazdjian's results for the scalar product derived from $j_{\rm int}^{\mu \nu}$ according to \eqref{eq:scalarprod} are as follows. The above-mentioned choice of Green's functions leads to the following expression\footnote{We have adopted our notation conventions. Besides, the range of integration is corrected according to the remark in sec.\@ \ref{sec:multitimecurrents} below eq.\@ \eqref{eq:scalarprod_sazdjian} such that the integration over $x_1$ and $x_2$ is over the same equal-time hypersurface instead of two different ones.} for a scalar product, for two eigenfunctions\footnote{For more general wave functions such as in \eqref{eq:momentumeigenfunctionsuperposition}, the definition of $\langle \cdot, \cdot \rangle_{\Sigma_t}$ can be extended by linearity.} $\psi_P,\psi_{P'}$ of $\hat{P}$ and the special case of $\Sigma = \Sigma_t$, i.e.\@ an equal-time hypersurface with normal covector field $n \equiv (1,0,0,0)$ \cite[eq.\@ (5.11)]{sazdjian_scalarprod}:
\begin{align}
 \langle \psi_P, \psi_{P'}\rangle_{\Sigma_t} ~&:=~ \int_{\Sigma_t \times \Sigma_t} d \sigma(x_1) d \sigma(x_2) \, j_{\rm int}^{\mu \nu}[\psi_P,\psi_{P'}](x_1,x_2) n_\mu(x_1) n_\nu(x_2)\nonumber\\
 &\,=~ \lim_{\varepsilon \rightarrow 0} \int d^3 \mathbf{X}\, d^3 \mathbf{x} ~ \overline{\psi}_P \left[ \gamma_1^0 \gamma_2^0 - V^*_{P'} \gamma_1^0 \gamma_2^0 V_P \right.\nonumber\\
 &~~~~~~~~~~~\left.+ ~({P^0}' + P^0) \frac{V_{P' + i\varepsilon n} - V_{P - i\varepsilon n}}{{P^0}' - P^0 + 2 i \varepsilon} \right] \psi_{P'}
 \label{eq:sazdjiansscalarprod}
\end{align}
where $(\cdot)^*$ denotes complex conjugation (without transposition). The limit $\varepsilon \rightarrow 0^+$ comes from the definition of $G_A(x)$ by its Fourier transform
\begin{equation}
 G_A(x) ~=~ \lim_{\varepsilon \rightarrow 0^+} \int \frac{d^4 k}{(2 \pi)^4} \frac{e^{-ik\cdot x}}{k^2 - 2ik^0 \varepsilon}.
\end{equation}
Let $\psi_P = e^{-i P \cdot X} \phi_1(x)$, $\psi_{P'} = e^{-i P' \cdot X}\phi_2(x)$. Then for $P^2 = (P')^2$, Sazdjian obtains from eq.\@ \eqref{eq:sazdjiansscalarprod} \cite[eq.\@ (5.12)]{sazdjian_scalarprod}:
\begin{equation}
 \langle \psi_P, \psi_{P'}\rangle_{\Sigma_t} ~=~ (2\pi)^3 \delta^{(3)}(\mathbf{p}-\mathbf{p}') \int d^3 \mathbf{x} \, \overline{\phi}_1(x) \left[ \gamma_1^0 \gamma_2^0 - V_P \gamma_1^0 \gamma_2^0 V_P + 4 (P^0)^2 \frac{\partial V_P}{\partial (P^2)}\right] \phi_2(x)
 \label{eq:normsazdjian}
\end{equation}
where $x = (x^0=0,\mathbf{x})$. We note that because of the delta function one can use $P = P'$ everywhere inside the integral.
\begin{remark}
 Ignoring the delta function, for $P = P'$, eq.\@ \eqref{eq:normsazdjian} should yield the square of a norm. It therefore has to be positive. However, in general (i.e.\@ independently of $V_P$) only the first term in the square brackets of eq.\@ \eqref{eq:normsazdjian}, which corresponds to the usual expression $\int \psi^\D \psi = \int \overline{\psi} \gamma_1^0 \gamma_2^0 \psi $ in the Dirac case, yields a positive contribution. Thus, as recognized by Sazdjian \cite[p. 1631]{sazdjian_scalarprod}\footnote{For clarity, notation and references in the quote are adapted to our conventions, without changes in content.}:
\begin{quotation}
 \noindent If the potential $V$ is explicitly independent of $P^2$ in the c.m.\@ frame, the expression of the norm \eqref{eq:normsazdjian} shows that its kernel still depends on $V$. This implies that $V$ must satisfy some inequality conditions to guarantee the positivity of the norm. This question was examined in more detail in Ref. \cite[sec.\@ VII A]{sazdjian_2bd}.
\end{quotation}
 In Ref. \cite[sec.\@ VII A]{sazdjian_2bd}, it is suggested to make a wave function transformation \cite[eq.\@ (7.1)]{sazdjian_2bd} which would map the norm given by eq.\@ \eqref{eq:normsazdjian} to the free $\int |\psi|^2$ norm. However, this is only possible for operators $\hat{V}$'s for which $V_P$ does not depend on $P^2$ in the c.m.\@ frame \cite[p. 3423]{sazdjian_2bd} and if in addition the following condition \cite[eq.\@ (7.6)]{sazdjian_2bd} is satisfied:
 \begin{equation}
  \tfrac{1}{4}\, {\rm Tr} \, (\gamma_1 \cdot \tfrac{\hat{P}}{\sqrt{\hat{P}^2}} ~\gamma_2 \cdot \tfrac{\hat{P}}{\sqrt{\hat{P}^2}} \hat{V}) ~ < ~ 1.
  \label{eq:tracecond}
 \end{equation}
 Presumably, the trace is to be taken over the spin components of $\hat{V}$. The question whether the independence of $V_P$ of $P^2$ in the c.m.\@ frame is a reasonable assumption is not clarified in \cite{sazdjian_2bd}. The more recent article \cite[eqs.\@ (A4), (A9)]{mourad_sazdjian} even seems to show the contrary.\\
 However, set aside this confusing point, a much more basic question is left open. Because $V = V_P$ and because $P$ is a property of $\psi_P$, i.e.\@ of the wave function, it is unclear how one should regard conditions that lead to the positivity of \eqref{eq:normsazdjian}:
 \begin{enumerate}
  \item as conditions on the space of admissible wave functions, or
  \item as conditions on the operators $\hat{V}$, given their domain?
 \end{enumerate}
 These questions are discussed in none of the references \cite{sazdjian_2bd,mourad_sazdjian,sazdjian_scalarprod} for the 2BD case.
 \end{remark}
 In the following subsections we analyze the consequences of these two possibilities (see sec.\@ \ref{sec:restrictedfunctions} for possibility 1 and sec.\@ \ref{sec:restrictedpotentials} for possibility 2).

\subsection{Can the positivity of norm and currents be guaranteed by restriction of the function space?} \label{sec:restrictedfunctions}
Recall from section \ref{sec:multitimecurrents} that the hope is to be able to regard the 2BD equations as defining an evolution map\footnote{Note that this question is independent of the fact that the operators $D_i$ appearing in the 2BD equations are not operators on Hilbert space.} for any pair of space-like hypersurfaces $\Sigma, \Sigma'$, i.e.\@ 
\begin{equation}
 U_{\Sigma \rightarrow \Sigma'} :~ \mathcal{H}_\Sigma^{(2)} \rightarrow \mathcal{H}_{\Sigma'}^{(2)},~~~\psi_\Sigma \mapsto \psi_{\Sigma'}
 \label{eq:timeevolmap}
\end{equation}
with $\psi_\Sigma(q) = \psi_{\Sigma'}(q)$ if $q \in \Sigma \cap \Sigma'$. $U_{\Sigma \rightarrow \Sigma'}$ should be unitary in the scalar product defined by $j_{\rm int}$ according to eq.\@ \eqref{eq:scalarprod}.\\
However, as we saw above, this construction does not yield a scalar product on $\mathcal{H}_\Sigma^{(2)}$ because it is in general not positive definite. Thus, we define
\begin{equation}
 \mathcal{H}^{\rm pos}_{\Sigma} ~:=~ \{ \phi \in \mathcal{H}_\Sigma^{(2)} : \langle \phi , \phi \rangle_\Sigma < \infty \wedge \langle \phi , \phi \rangle_\Sigma > 0 \} \cup \{ 0\}
 \label{eq:restrictedspace}
\end{equation}
as the subspaces of $\mathcal{H}_\Sigma^{(2)}$ for which $\langle \cdot , \cdot \rangle_\Sigma$ is actually positive-definite. The question is: \textit{does $\mathcal{H}^{\rm pos}_{\Sigma}$ define an acceptable space of functions?}\\

\noindent To decide on this question, consider the following points:
\begin{enumerate}
 \item It is not clear anymore that $\mathcal{H}^{\rm pos}_{\Sigma}$ contains all physically relevant functions. One can see this e.g.\@ from \eqref{eq:momentumeigenfunctionsuperposition} and \eqref{eq:normsazdjian}. Any reasonable quantum mechanical theory which describes matter should be able to describe localized wave packets. To construct these wave packets, one in general requires all Fourier modes $\psi_P$ as in \eqref{eq:momentumeigenfunctionsuperposition}. However, for a general $\hat{V}$, the requirement of positivity of \eqref{eq:normsazdjian} implies conditions on the $P$'s such that some are not admitted in the construction of wave packets. Furthermore, these conditions are mathematically quite involved and do not serve a clear physical purpose.
 \item The $\mathcal{H}^{\rm pos}_{\Sigma}$ do not, in general, define Hilbert spaces. Completeness may be violated. Even worse, the $\mathcal{H}^{\rm pos}_{\Sigma}$ may not even be vector spaces. Consequently, the mathematical structures on which both the usual quantum formalism is built up, break down, including the self-adjoint operator observables as well as the standard approach to define the time evolution. Of course, one may consider to further reduce the admissible functions by replacing $\mathcal{H}^{\rm pos}$ with some Hilbert space $\mathcal{H}^*$ contained in it. In fact, a similar route was suggested by Sazdjian \cite[p. 1624]{sazdjian_scalarprod}. This, however, further strengthens the criticism of point 1 and still leaves open the question if the usual mathematical structures can be defined.
\end{enumerate}	

\paragraph{A simple analogy:} To appreciate the problems that accompany the restricted function spaces $\mathcal{H}^{\rm pos}_{\Sigma}$ and $\mathcal{H}^*$, consider the following example. Let the Hilbert space of our theory be given by $\C^2$ with ``scalar product''
\begin{equation}
 \langle v, w\rangle_A~:=~ v^\D A w,~~~~~{\rm where}~~~A = \left( \begin{array}{cc}
                                                       1& 0\\ 0&-1
                                                      \end{array} \right).
 \label{eq:analogyscalarprod}
\end{equation}
Of course, $\langle \cdot , \cdot \rangle_A$ does not define a scalar product on $\C^2$. So in analogy with \eqref{eq:restrictedspace} we define
\begin{equation}
 \mathcal{H}^{\rm pos} ~:=~ \{ v \in \C^2 : v^\D A v > 0\} \cup \{ 0\}
 \label{eq:analogyhpos}
\end{equation}
as the subset on which $\langle \cdot , \cdot \rangle_A$ actually is a scalar product. We note that e.g.\@ $(0,1)$ (which may be a physically relevant vector to represent a spin state) is not contained in $\mathcal{H}^{\rm pos}$ (cf. point 1).\\
Moreover, $\mathcal{H}^{\rm pos}$ is not a vector space, because the for $v_1 = (1, \tfrac{1}{2}) \in \mathcal{H}^{\rm pos}$ and $v_2 = (-1, \tfrac{1}{2}) \in \mathcal{H}^{\rm pos}$, the sum $v_1 + v_2 = (0,1)$ is not an element of $\mathcal{H}^{\rm pos}$. Furthermore, it is also not complete, as the following example illustrates: Consider the sequence given by $v_n = (1, 1-1/n)$. We have: $\langle v_n, v_n \rangle_A = 1- (1- 1/n)^2 > 0$ and thus $v_n \in \mathcal{H}^{\rm pos}$. However, the limit $v = (1,1)$ has norm zero, i.e.\@ $v \not\in \mathcal{H}^{\rm pos}$. These problems are analogous to point 2 from above. Note that they can be overcome by defining even smaller Hilbert spaces $\mathcal{H}^*$ as the span of $(1,0)$ instead. ($\mathcal{H}^*$ is a complete vector space for which $\langle \cdot, \cdot \rangle_A$ defines a scalar product). However, even more physically interesting vectors get lost this way.\\
To extend the analogy, suppose that the ``wave equation'' of our theory is
\begin{equation}
 i \frac{d}{d t} u ~=~ B u
\label{eq:toywaveeq}
\end{equation}
where $B$ is the following $2 \times 2$ matrix
\begin{equation}
 B ~=~ \left( \begin{array}{cc}
               0&i\\ -i&0
               \end{array} \right).
\end{equation}
The matrix $B$ is self-adjoint with respect to the canonical scalar product on $\C^2$ but not with respect to $\langle \cdot, \cdot \rangle_A$. Thus, it defines a time evolution on $\C^2$ but not necessarily on $\mathcal{H}^{\rm pos}$. Let us analyze the consequences. Given $u(t = 0) \equiv u_0$, we have
\begin{equation}
  u(t)~=~\exp(-iBt)u_0 ~=~ (\id_2 \, \cos t~ -i B \, \sin t) u_0.
 \label{eq:timevolu}
\end{equation}
Let $u_0 = (1,0) \in \mathcal{H}^{\rm pos} \cap \mathcal{H}^*$. Then $u(t) = (\cos t, - \sin t)$ which is in general not an element of $\mathcal{H}^{\rm pos}$ (neither of $\mathcal{H}^*$).\\
One may try admitting only initial data $u_0 = (a,b) \in \mathcal{H}^{\rm pos}$ for which also $u(t) \in \mathcal{H}^{\rm pos}~\forall t$. Then $|a| > |b|$. We have $u(t) = (a \cos t + b \sin t, - a \sin t + b \cos t)$ and therefore
\begin{equation}
 u^\D(t) A  u(t) ~=~ (|a|^2 - |b|^2)(\cos^2 t - \sin^2 t) + 4 \, {\rm Re} \, (a^* b) \cos t \sin t.
 \label{eq:normu}
\end{equation}
We ask: Do $a,b \in \C$ with $|a|>|b|$ exist which make this expression positive for every $t$? For an answer, consider \eqref{eq:normu} for (i) $t = \pi/4$ and (ii) $t = 3\pi/4$. In case (i), we have $\sin t = \cos t = 1/\sqrt{2}$ and obtain as a condition that ${\rm Re} \, (a^* b) >0$. In case (ii), $\sin t = -\cos t = 1/\sqrt{2}$ and we obtain the condition ${\rm Re} \, (a^* b) <0$, in contradiction with (i). We conclude that the restriction to $\mathcal{H}^{\rm pos}$ is not in any way consistent with the given time evolution \eqref{eq:toywaveeq} (neither is the restriction to $\mathcal{H}^*$). This illustrates the problems with defining the time evolution from point 2 above.

\paragraph{Comparison with Klein-Gordon theory:} If the above analogy extends to the case of the 2BD equations, the logical consequence is to reject restrictions on the function space. However, in view of previous claims about the consistency of a Hilbert space picture for interacting two-body Klein-Gordon (KG) equations in \cite{rizov_tensor_currents,longhi_lusanna_86,sazdjian_scalarprod}, one may wonder if points 1,2 are actually as severe as they seem to be.\\
These sources (especially \cite[sec.\@ III]{sazdjian_scalarprod}) show the following. In the KG case, one can identify potentials such that the scalar product given by the conserved tensor currents of the theory according to \eqref{eq:scalarprod} is positive definite on a subspace $\mathcal{H}^*$ of the Hilbert space $\mathcal{H}_{\Sigma}$ of the theory. Then $\mathcal{H}^*$ is again a Hilbert space, corresponding to one of four possible choices of the sign of eigenvalues of the operators $p_1 \cdot \hat{P}/\hat{P}^2$ and $p_2 \cdot \hat{P}/\hat{P}^2$. One may thus hope that the problems of point 2 do not appear.\\
However, this approach seems to disregard problem 1: the subspace $\mathcal{H}^*$ does not contain all physically relevant functions. One cannot, for example, represent localized wave packets by wave functions in $\mathcal{H}^*$. To do so would require basis vectors from all of $\mathcal{H}_\Sigma$. Furthermore, problems with the self-adjointness of operator observables may occur (see also \cite[p. 66]{rizov_tensor_currents}). A completely analogous situation is encountered in free one-particle KG theory \cite[chap. 3]{schweber}. In this case, one draws the logical consequence that the KG equation theory cannot be considered a self-contained one-particle theory. By the same arguments, one also has to reject the approach via $\mathcal{H}^*$ towards interacting two-body KG theory.

\paragraph{Conclusion:} One may wonder whether or not the situation for KG theory has any significance for the 2BD theory. As remarked after the discussion of the meaning of $x_\perp$ in sec.\@ \ref{sec:implicationskb}, the square of a the 2BD equations does in general not yield an interacting KG equation. Moreover, recalling the quote at the end of section \ref{sec:previousresults}, the implications of the two-body KG theory on the 2BD theory are limited. For the 2BD case, the ``scalar product'' is not positive definite even if the norm is independent of $P^2$ in the c.m.\@ frame (see also \cite[p. 1627-28, 1631]{sazdjian_2bd}).  Furthermore, we note that the approach in the KG case involves both restrictions on the potential operator $\hat{V}$ as well as restrictions on the function space. The restrictions on the function space turned out to be inacceptable whereas there is no reason to reject restrictions on the potentials as long as they include the ones used in applications. In 
 view of this situation, 
together with points 1-2 (as illustrated by the analogy), we conclude that restrictions on the function space are also inacceptable for the 2BD theory\footnote{One may even hope that the situation is better in the 2BD theory in the sense 
that the free Dirac current is positive definite, as opposed to the free KG current.}. The question if, on the other hand, there exist sensible restrictions on the potentials such that the scalar product is positive definite is the subject of the next section.

\subsection{Do special operators $\hat{V}$ exist for which scalar product and currents are always positive definite?} \label{sec:restrictedpotentials}
Consider eq.\@ \eqref{eq:normsazdjian} ``in the c.m.\@ frame'', i.e.\@ for $P = (P^0,0,0,0)$. Then $\hat{x}_\perp$ acts as the multiplication operator with the spatial part $\mathbf{x}$ of the relative coordinate. Demanding that $\langle \psi_P, \psi_P\rangle_{\Sigma_t}$ be positive for all eigenfunctions $\psi_P$ of $\hat{P}$, we obtain the following condition for $V_P$:
\begin{equation}
 \overline{\phi}(\mathbf{x}) \left[ \gamma_1^0 \gamma_2^0 - V_P \gamma_1^0 \gamma_2^0 V_P + 4 (P^0)^2 \frac{\partial V_P}{\partial (P^2)}\right] \phi(\mathbf{x}) ~ \geq 0.
\label{eq:condv}
\end{equation}
This condition should be satisfied for a reasonably general class of functions $\phi$, e.g.\@ for all $\phi \in L^2(\R^3) \otimes \C^{16}$. We note the following points:
\begin{enumerate}
 \item $V_P$, which depends on $\hat{P}$ via the quantities $x_\perp = (0,\mathbf{x})$ and $P^2 = (P^0)^2$ in the c.m.\@ frame, has to be bounded\footnote{``Bounded'' in the context means that the absolute values of the eigenvalues of $V_P = V(x_\perp,P)$ are bounded.} with respect to (a) $P^2$ and (b) $x_\perp^2 = - \mathbf{x}^2$. (a) is easy to achieve, e.g.\@ in the case that $V$ does not depend on $(P^0)^2$ in the c.m.\@ frame. (b) is a real restriction. We shall see the consequences below.
 \item Do solutions $V_P$ of eq.\@ \eqref{eq:condv} exist? To answer this question, consider the class of scalar functions $V_P \equiv f(-x_\perp^2)$. In the c.m.\@ frame they take the form $f(\mathbf{x}^2)$ which is independent of $(P^0)^2$. Thus, the term $4 (P^0)^2 \frac{\partial V_P}{\partial (P^2)}$ in eq.\@ \eqref{eq:condv} vanishes. Making use of the fact that $f(-\mathbf{x}^2)$ is real-valued as a consequence of eq.\@ \eqref{eq:hermiticitycond} in the scalar case, condition \eqref{eq:condv} reduces to:
 \begin{align}
  \phi^\D(\mathbf{x}) \left[ \id - |f(\mathbf{x})|^2\right] \phi(\mathbf{x}) ~ &\geq~ 0~~\forall \phi \in L^2(\R^3) \otimes \C^{16},~\forall \mathbf{x} \in \R^3 \nonumber\\
  \Leftrightarrow ~~~|f(\mathbf{x}^2)| ~& \leq ~ 1~\forall \mathbf{x} \in \R^3.
  \label{eq:condvscalar}
 \end{align}
\end{enumerate}
 \textit{Thus, we conclude that there do indeed exist special operators,} e.g.\@ $\hat{V} \equiv f(-\hat{x}_\perp^2)$ with $|f(y)| < 1~\forall y \in \R$, \textit{for which the scalar product is positive definite on a general function space,} e.g.\@ for superpositions of eigenfunctions of $\hat{P}$ \eqref{eq:momentumeigenfunctionsuperposition} with suitable drop-off conditions\footnote{Note that because of the form of the kernel of the scalar product \eqref{eq:normsazdjian} the drop-off conditions may become modified as compared to the case $\hat{V} \equiv 0$. If this should turn out problematic, one could easily avoid the situation by demanding that $V_P$ falls off with $|x_\perp^2| \rightarrow \infty$ sufficiently fast.} for $|x_\perp^2| \rightarrow \infty$. Given any smooth and real-valued function $g(y)$, such a function $f$ can be constructed as
 \begin{equation}
  f(y)~:=~\tanh g(y).
 \end{equation}
  One may, however, ask: \textit{are these restrictions on $\hat{V}$ physically reasonable\footnote{Note that when comparing such a bounded $\hat{V}$ with, say, a Coulomb potential (which is unbounded), one may have to take into account a wave function transformation (see \cite{mourad_sazdjian} and the remark below eq.\@ \eqref{eq:formdi}).}?} \\
  We try to answer this question by comparison with realistic potentials derived from quantum field theory in \cite[appendix A]{mourad_sazdjian}. One such possibility for scalar interactions in lowest order perturbation theory is \cite[eqs.\@ (A4), (2.17), (2.20)]{mourad_sazdjian}:
 \begin{equation}
  \hat{V}_1 ~:=~ \tanh \left[ -\frac{1}{2\sqrt{\hat{P}^2}} \frac{g_1 g_2}{4 \pi} \frac{\exp\left(-\mu \sqrt{-\hat{x}_\perp^2}\right)}{\sqrt{{-\hat{x}_\perp^2}}} \right],
  \label{eq:v1}
 \end{equation}
 where $g_1, g_2 \in \R$, $\mu > 0$. The question is: \textit{does $\hat{V}_1$ fulfil the positivity condition \eqref{eq:condv}?}\\
 We first note that $V_{1,P}$ (i.e.\@ $\hat{V}_1$ where $\hat{P}$ is replaced by an eigenvalue $P$) does indeed explicitly depend on $P^2$ even for $P = (P^0,0,0,0)$. This feature is shared with other possible potentials derived from QFT (see\footnote{Note that also for Crater and Van Alstine's form of the equations the potentials explicitly depend on $\sqrt{P^2}$, called $w$ in the references (see e.g. \cite[appendix A]{crater_applications}).} \cite[appendix A]{mourad_sazdjian}). Thus, we can neither use the simplified condition \eqref{eq:condvscalar} and nor the before-mentioned condition \eqref{eq:tracecond} of Sazdjian.\\
 Let us evaluate condition \eqref{eq:condv} for $V_{1,P}$ for eigenfunctions of $\hat{P}$ and in the c.m.\@ frame. Then: $P^2 = (P^0)^2$ and $\sqrt{-x_\perp^2} = |\mathbf{x}|$. We have:
 \begin{equation}
  \frac{\partial V_{1,P}}{\partial (P^0)^2} ~=~ \frac{1}{4} |P^0|^{-3} \frac{g_1 g_2}{4 \pi} \frac{e^{-\mu |\mathbf{x}|}}{|\mathbf{x}|} \frac{1}{\cosh^2 \left[ -\frac{1}{2|P^0|} \frac{g_1 g_2}{4 \pi} \frac{e^{-\mu |\mathbf{x}|}}{|\mathbf{x}|} \right]}.
  \label{eq:dv1p}
 \end{equation}
 Let
 \begin{equation}
  y ~:=~ \frac{1}{2|P^0|} \frac{g_1 g_2}{4 \pi} \frac{e^{-\mu |\mathbf{x}|}}{|\mathbf{x}|}.
  \label{eq:y}
 \end{equation}
 Evidently, $y > 0$. Then eq.\@ \eqref{eq:condv} becomes
 \begin{equation}
  \phi^\D \left[ \id (1- \tanh^2(-y)) + 2 \gamma_1^0 \gamma_2^0 \, \frac{y}{\cosh^2(-y)}\right] \phi ~ \geq 0~~ \forall y > 0,~\forall \phi \in \C^{16}.
  \label{eq:condv2}
 \end{equation}
 Note that $\gamma_j^0$ has eigenvalues $\pm 1$ for each $j$. Thus, eq.\@ \eqref{eq:condv2} yields the two conditions
 \begin{equation}
  1 - \tanh^2(-y) \pm \frac{2y}{\cosh^2(-y)}~\geq 0~~\forall y > 0.
  \label{eq:condv3}
 \end{equation}
 However, for ``$-$'', the function $h(y) := 1 - \tanh^2(-y) - \frac{2y}{\cosh^2(-y)}$ is negative for $y > \tfrac{1}{2}$, corresponding to $|\mathbf{x}|e^{\mu |\mathbf{x}|} < \frac{1}{|P^0|} \frac{g_1 g_2}{4 \pi}$. Consequently, wave functions with internal part $\phi(\mathbf{x})$ with support concentrated around $|\mathbf{x}| = 0$ have negative ``norm'' and ``probability density''.

\paragraph{Comparison with the norm used by Crater and Van Alstine:}
Building on Sazdjian's work \cite{sazdjian_scalarprod}, Crater and Van Alstine also considered the question of an adequate norm \cite[p. 9]{2bd_hyperbolic}. Their derivation of the norm is based on the following wave function transformation between the wave function $\psi$ of eq.\@  \eqref{eq:2bdsazdjian} and the wave function $\tilde{\psi}$ appearing in the so-called ``hyperbolic form'' of of their equations \cite[eqs.\@ (52), (53)]{2bd_hyperbolic}:
\begin{equation}
 \psi~=~ \cosh(\Delta) \tilde{\psi}
\label{eq:wavefntrafo}
\end{equation}
where
\begin{equation}
 \Delta ~=~ \tanh^{-1}(\hat{V}).
 \label{eq:potentialtrafo}
\end{equation}
We note that this transformation is not a simple mathematical object because it evidently depends on the operator $\hat{V}$. We continue with the assumption that it does indeed exist (which may yield further conditions on the potentials or on the function space) and analyze the consequences for the norm.\\
Employing the transformation \eqref{eq:wavefntrafo} for eq.\@ \eqref{eq:normsazdjian}, Crater and Van Alstine obtain (see \cite{2bd_hyperbolic} and \cite[appendix B]{crater_norm}; the result is adapted to our notation):
\begin{equation}
 \langle \tilde{\psi}_P, \tilde{\psi}_{P'}\rangle_{\Sigma_t} ~=~ (2\pi)^3 \delta^{(3)}(\mathbf{p}-\mathbf{p}') \int d^3 \mathbf{x} \, \tilde{\phi}^\D_1(x) \left[ \id - 4 P^2 \gamma_1^0 \gamma_2^0 \frac{\partial \Delta_P}{\partial (P^2)}\right] \tilde{\phi}_2(x)
 \label{eq:cratersnorm}
\end{equation}
where the $\tilde{\phi}_i$ are defined analogously to the $\phi_i$ in eq.\@ \eqref{eq:normsazdjian} and $\Delta_P$ is the operator $\Delta$ with $\hat{P}$ replaced by its eigenvalue $P$. The symbol $\tilde{(\cdot)}$ indicates that the  wave function transformation \eqref{eq:wavefntrafo} has been made.\\
Considering eq.\@ \eqref{eq:cratersnorm}, we note that the expression for the norm (i.e.\@ for $P = P'$) reduces to the usual $\int |\psi|^2$-expression \textit{for energy-independent potentials} and is then positive without further restriction on the potentials. However, as evident from both \cite[appendix A]{mourad_sazdjian} and \cite[appendix A]{crater_applications}, \textit{realistic choices of the potentials explicitly depend on the energy} $\sqrt{P^2}$. Thus, equivalent restrictions on the potentials as given by condition \eqref{eq:condv} also appear following Crater's and Van Alstine's approach.  This, of course, has to be the case if the wave function transformation is to yield an equivalence between the 2BD equations of Sazdjian \eqref{eq:2bdsazdjian} and the 2BD equations of Crater and Van Alstine.\\
More precisely, one can see from eq.\@ \eqref{eq:cratersnorm} that the following condition has to be satisfied by $\Delta$:
\begin{equation}
 \tilde{\phi}^\D(\mathbf{x}) \left[\id - 4 P^2 \gamma_1^0 \gamma_2^0 \frac{\partial \Delta_P}{\partial (P^2)}\right] \tilde{\phi}(\mathbf{x}) ~ \geq 0
 \label{eq:condvcrater}
\end{equation}
for all $\tilde{\phi} \in L^2(\R^3) \otimes \C^{16}$.\\
We now evaluate this condition for a realistic choice of $\Delta$, corresponding to $\hat{V}_1$ from above (see eq.\@ \eqref{eq:v1}, \cite[eq.\@ (A4)]{mourad_sazdjian}). Then:
\begin{equation}
 \Delta_1 ~=~ \tanh^{-1}(\hat{V}_1)~=~ -\frac{1}{2\sqrt{\hat{P}^2}} \frac{g_1 g_2}{4 \pi} \frac{\exp\left(-\mu \sqrt{-\hat{x}_\perp^2}\right)}{\sqrt{{-\hat{x}_\perp^2}}}.
 \label{eq:delta_1}
\end{equation}
After a short and elementary calculation similar to the one leading from eq.\@ \eqref{eq:condv} to eq.\@ \eqref{eq:condv3}, condition \eqref{eq:condvcrater} reduces to
\begin{equation}
 |\mathbf{x}| e^{\mu |\mathbf{x}|} ~ <~ \frac{g_1 g_2}{4 \pi}
 \label{eq:resultcondvcrater}
\end{equation}
which is the same condition as before, with the same consequences.

\paragraph{Conclusion:} It is in principle possible to guarantee the positive definiteness of the scalar product by special choices for the potential operator $\hat{V}$. This is particularly easy to achieve in the scalar case and for $\hat{V}$'s which are  independent of $\hat{P}^2$. Realistic choices for $\hat{V}$ such as $\hat{V}_1$ from eq.\@ \eqref{eq:v1}, however, are not independent of $\hat{P}^2$. This has the consequence that $\hat{V}_1$ does not satisfy the condition \eqref{eq:condv} for positive definiteness of the scalar product and of the probability density. One may suspect that other realistic choices for $\hat{V}$ might suffer from the same problem. Therefore, they might not lead to a self-contained quantum mechanical two-particle theory which can possibly make statistical predictions in its own right.

\section{Gauge invariance} \label{sec:gaugeinvariance}
In this section, we briefly comment on how the unusual mathematical structure of the 2BD equations influences the notion of gauge invariance.\\
According to the view put forward in sec.\@ \ref{sec:multitimecurrents}, one should regard the tensor current $j^{\mu \nu}[\psi,\psi]$, not the wave function $\psi$, as the physical object. Transformations $\psi$ which leave $j$ invariant are considered pure gauge. In the case of free multi-time Dirac equations, $j^{\mu \nu}[\psi,\psi] = \overline{\psi}\gamma_1^\mu \gamma_2^\nu \psi$ and the gauge transformations are given by (see also \cite{nogo_potentials}):
\begin{equation}
 \psi(x_1,x_2)~\mapsto~e^{-i \theta(x_1,x_2)} \psi(x_1,x_2).
 \label{eq:gaugetrafo}
\end{equation}
In the 2BD case, however, the tensor currents in the 2BD case have to be modified (see sec.\@ \ref{sec:previousresults}). The possible replacements are momentum-dependent, i.e.\@ their form depends on the wave function itself. Consequently, the form of a gauge transformation changes. In particular, the standard transformations \eqref{eq:gaugetrafo} cannot in general be considered gauge transformations anymore, because e.g.\@ $\psi \mapsto e^{-i P \cdot X} \psi$ may change the eigenvalue $P$ of $\hat{P}$ and $P$ in turn is crucial for the form of $j_{\rm int}$. The class of gauge transformations is thus reduced to transformations
\begin{equation}
 \theta(x_1,x_2) ~\equiv~ \tilde{\theta}(x)
 \label{eq:restrictedgaugetrafo}
\end{equation}
which do not involve the coordinate $X$ on which $\hat{P}$ acts. Note, however, that the general gauge transformations \eqref{eq:gaugetrafo} may introduce terms into the multi-time equations \eqref{eq:generalform2bd} which are not Poincaré invariant. This is not possible using only the restricted class of gauge transformations \eqref{eq:restrictedgaugetrafo}.

\section{Discussion} \label{sec:discussion}

In this paper, we critically reviewed various questions around the 2BD equations, rectifying critical issues where needed and pointing out unsolved problems. First, we presented in detail how the 2BD equations connect to the important work of Dirac, Bloch, Tomonaga and Schwinger on the multi-time formalism, as well as to recent developments. At this, we presented a general framework how to understand multi-time equations both mathematically and physically. The latter aspect led to the requirement of conserved tensor currents with a positive component which can play the role of a probability density. The importance of this requirement cannot be stressed enough because it constitutes the link between the mathematical formalism and the statistical outcomes of experiments. Moreover, we hinted at the connection to realistic relativistic quantum theories.\\
Starting from the general framework, we showed how the 2BD equations avoid a recent no-go theorem which, contrary to older well-known results such as the so-called ``no interaction theorem'', is directly applicable to multi-time Dirac equations, not only up to quantization. The relevant mechanism for this is the inclusion of arbitrary powers of the total momentum operator in the interaction terms. This allows to satisfy a necessary compatibility condition by assuming a certain relation between the interaction terms. The compatibility condition then implies the appearance of (an operator version of) the variable $x_\perp$. In classical mechanics, $x_\perp$ is the covariantization of the spatial distance in the center of momentum frame. The subtle arguments why the use of such a covariantization can be considered in agreement with relativistic physics were carefully discussed.\\
We continued by focusing on a certain, still fairly general class of the 2BD equations for particle-antiparticle pairs which was first suggested by Sazdjian. Here, we pointed out the most important mathematical questions (which have not recieved sufficient attention in the literature) and proposed a preliminary mathematical understanding of the 2BD equations for superpositions of total momentum eigenfunctions. The question of the existence and uniqueness of solutions of the 2BD equations, however, remains open.\\
The main part of the paper was devoted to the above-mentioned critical question of whether conserved and positive definite tensor currents exist. While the free Dirac currents are not conserved, there do exist possible replacements. These replacements are not unique and there is no general argument why for any of them the currents should be positive definite for arbitrary potentials and on a general function space. This situation motivated the question whether the currents can be rendered positive definite by restricting the function space or the admitted class of potentials. Our analysis started out from a previous one by Sazdjian which was, however, shown incomplete. In particular, Sazdjian did not discuss the question if further restrictions to render the currents positive definite are to be regarded as restrictions of the function space or of the potentials.\\
These two aspects were the topic of the next two sections. First, we showed in detail that restrictions of the function space are not acceptable. The reason is that, roughly speaking, Fourier modes are excluded which are necessary for the representation of localized wave packets, for the self-adjointness of operators and for the usual way of defining the time evolution. We also commented on existing literature on similar questions, finding that a comprehensive analysis has never been done before for the Dirac case (the Klein-Gordon case anyway not leading to positive-definite currents).\\
Second, we analyzed the implications of the requirement of positivity of the currents on the allowed form of the potentials, given a sufficiently general function space. The results were twofold: on the one hand, we found that it is indeed possible to identify a general class of potentials with the desired property. On the other hand, potentials which were suggested as physically accurate in the literature may in fact violate the requirements for positive definite currents. It should be emphasized that in any case the form of the probability density changes as a consequence of the fact that the form of the currents depends on the chosen potential. This is of relevance also for applications, e.g.\@ concerning transition rates. (For spectra, on the other hand, the form of the probability density is unimportant.)\\

As there are several possible motivations to study the 2BD equations, the implications of the results on restrictions for the admissible potentials can be regarded in different ways. On the one hand, for applications where realistic potentials are required, doubts are raised that phenomenological calculations of meson spectra based on the 2BD equations like in \cite{crater_meson_spectroscopy,crater_applications,mourad_sazdjian,2bd_exactsolution,crater_norm} do have a theoretical justification. To resolve the doubts would require to check the positivity condition \eqref{eq:condv} (or, equivalently, \eqref{eq:condvcrater}) for the potential used. This, however, is not done in the literature, nor does an awareness of the problem seem to exist. In fact, we found that for a simple choice of the potential which was derived by Sazdjian from the Bethe-Salpeter equation, it is violated. The fact that this topic has not received attention before is somewhat surprising, considering that one of 
 Crater's and Van Alstine's main 
motivations to introduce the first version of 
the 2BD equations was that the Bethe-Salpeter equation possesses negative-norm states -- and therefore does not have a clear physical interpretation \cite{2bd}.\\
Furthermore, assuming that physically realistic potentials could be found which also satisfy the positivity condition, the modified probability density as compared to the $|\psi|^2$-density seems unusual. It would therefore be interesting to subject the modified density to experimental tests, for example by determining transition rates. In this respect, Crater and coworkers have investigated decay rates of quarkonium and positronium into two photons\footnote{I am grateful to H. W. Crater for pointing this out to me.}, considering the effects of the modified norm \eqref{eq:cratersnorm} \cite{crater_norm}.
The theoretical results obtained compare well with other phenomenological approaches, but still lie outside of the error bars of the experimental data in all cases. These differences between theory and experiment are particularly interesting, considering that they appear at a place which is critical from a purely theoretical point of view: as we stressed in sec.\@ \ref{sec:previousresults}, the tensor current is not unique, requiring the choice of two Green's functions in eq.\@ \eqref{eq:jaddsolution}.  Note that the theoretical results for the mesonic spectra given in \cite{crater_norm}, which are independent of the exact form of the tensor current, fit much better with the experimental data. One could take these findings as a motivation to study the question whether modifying the potentials or (as is particularly interesting) making a different choice of the tensor current could improve the theoretical results.
However, before immediately drawing the consequence that such modifications are required, one should not forget that further (possibly critical) assumptions are involved in the process of calculating decay rates via the 2BD equations. This is obvious from the fact that the 2BD equations as a strict two-particle theory do not, by themselves, accomodate processes with variable particle numbers. A theoretical justification to nevertheless calculate decay rates using solutions of the 2BD equations therefore cannot be contained in the framework of the 2BD equations alone but has to come e.g.\@ from quantum field theory.\\ 
On the other hand, for foundational aspects in relativistic quantum theory, it seems remarkable that there do exist interaction terms for multi-time equations at all which satisfy the minimal requirements of Lorentz invariance and compatibility with a probabilistic meaning of the wave function. One class of these equations is given by\footnote{Note that a similarly looking class of 2BD equations was suggested by Crater and Van Alstine \cite[eqs.\@ (52), (53)]{2bd_hyperbolic}. Eq.\@ \eqref{eq:satisfactory2bdeqs} is a subclass of these equations for which the positivity of the scalar product and currents has been checked in sec.\@ \ref{sec:restrictedpotentials}. For the general class in \cite{2bd_hyperbolic}, positivity may be violated.}:
\begin{align}
&\left\{ \gamma_1 \cdot \hat{p}_1 - m_1 -(- \gamma_2 \cdot \hat{p}_2 + m_2) \tanh \left[ g(-\hat{x}_\perp^2) \right] \right\} \psi(x_1,x_2) ~=~ 0,\nonumber\\
 &\left\{ \gamma_2 \cdot \hat{p}_2 + m_2 +(\gamma_1 \cdot \hat{p}_1 + m_1)\tanh \left[ g(-\hat{x}_\perp^2) \right] \right\} \psi(x_1,x_2) ~=~ 0,
 \label{eq:satisfactory2bdeqs}
\end{align}
where $g(y)$ is an arbitrary smooth and real-valued function. The expression for the associated positive tensor current is rather lengthy and can be calculated via eqs.\@ \eqref{eq:f}, \eqref{eq:jaddsolution}. The corresponding scalar product, evaluated on equal-time hypersurfaces of a special frame, is given by \eqref{eq:normsazdjian}. As stressed above, the tensor current involved in this construction is not unique. Such a non-uniqueness of the currents is, however, not an uncommon situation in quantum physis. One can for example always add a term which is divergence-free -- and sometimes this is even appropriate. Moreover, the additional freedom in choosing a Green's function in eq.\@ \eqref{eq:jaddsolution} might (in more general situations than \eqref{eq:satisfactory2bdeqs}) help to reconcile experimental and theoretical results for decay rates.\\
Finally, one may wonder whether a similar approach as for the 2BD equations can be taken also for $N>2$ particles. However, appropriate wave equations in a closed form have never been found. This may be due to the fact that there does not exist a generalization of the variable $x_\perp$ for $N$ particles which allows to satisfy the necessary compatibility condition of the wave equations in a similar way as for two particles \cite{sazdjian_nbody,crater_wf}.

\subsection*{Acknowledgments}
I would like to thank Roderich Tumulka for helpful discussions, Lukas Nickel for constructive remarks on the manuscript and Horace Crater for a helpful and inspiring correspondence. Special thanks go to Detlef Dürr for many thorough discussions and insightful remarks, as well as for providing a different perspective. Financial support by the German National Academic Foundation is gratefully acknowleged.

\end{document}